\documentclass[12pt]{article}
\usepackage{a4,geometry,graphicx,epsfig,amsfonts,amsthm}
\usepackage{subfigure}
\newcommand{\be}{\begin{equation}}
\newcommand{\ee}{\end{equation}}
\newcommand{\bea}{\begin{eqnarray}}
\newcommand{\eea}{\end{eqnarray}}

\newcommand{\bkappa}{\mbox{\boldmath $\kappa$}}
\newcommand{\bk}{{\bf k}}
\newcommand{\bbe}{{\bf e}}

\newcommand{\bq}{{\bf q}}

\newcommand{\bb}{{\bf b}}

\def\lsim{\mathrel{\rlap{\lower4pt\hbox{\hskip1pt$\sim$}}
    \raise1pt\hbox{$<$}}}         

\def\gsim{\mathrel{\rlap{\lower4pt\hbox{\hskip1pt$\sim$}}
    \raise1pt\hbox{$>$}}}         

\begin{document}
%
~\vskip 2cm
\begin{center}
\bigskip
{\Large \bf \boldmath Central exclusive production of dijets\\ at 
hadronic colliders} \\
\bigskip
\bigskip
{\large J.R. Cudell$^{a,}$\footnote{jr.cudell@ulg.ac.be}, A.
Dechambre$^{a,}$\footnote{alice.dechambre@ulg.ac.be}, O. F.
Hern\'andez,$^{b,}$\footnote{oscarh@physics.mcgill.ca ; permanent
address: Marianopolis College, 4873 Westmount Ave., \\
\indent \hspace*{2.5mm}Montr\'eal, QC, Canada H3Y 1X9}
 I. P. Ivanov$^{a,c,}$\footnote{igor.ivanov@ulg.ac.be}}
~\vglue 1cm
{\small\it  $^{a}$ IFPA, D\'epartement AGO, Universit\'e de Li\`ege, Sart 
Tilman, 4000 Li\`ege, Belgium}

{\small\it  $^{b}$ Physics Dept., McGill University, 3600 University St., 
Montr\'eal, Qu\'ebec, Canada, H3A2T8}

{\small\it $^{c}$Sobolev Institute of Mathematics, Koptyug avenue 4, 
630090, Novosibirsk, Russia}

\end{center}
%
\begin{center}
\bigskip (\today)
\vskip0.5cm {\Large Abstract\\} \vskip3truemm
\parbox[t]{\textwidth}{In view of the recent diffractive dijet data from
CDF run II, we critically re-evaluate the standard approach to the
calculation of central production of dijets in quasi-elastic hadronic 
collisions. We find that the process is dominated by the
non-perturbative region, and that even perturbative ingredients,
such as the Sudakov form factor, are not under theoretical control.
Comparison with data allows us to fix some of the uncertainties.
Although we focus on dijets, our arguments apply to other high-mass
central systems, such as the Higgs boson.}
\end{center}
\thispagestyle{empty}
\newpage
\setcounter{page}{1}
%

\section*{Introduction}
\label{sec-introduction}

The CDF collaboration has recently published the measurement of the
cross section for exclusive dijet production~\cite{CDF}. These are
important data, as the dijet system reaches masses $M_{jj}$ of the order
of 130 GeV, {\it i.e.} the region of mass where the Higgs boson is
expected. Given the high centre-of-mass energy involved,
$\sqrt{s}\gg M_{jj}$, the process is entirely due to pomeron exchange.
Thus the dijets are produced by the same physical mechanism which
could produce the Higgs boson and other rare particles. The CDF data
provide a good opportunity to tune the calculations of quasi-elastic
diffractive processes which, as we shall show, are otherwise plagued
with severe uncertainties.

Almost twenty years ago, Sch\"afer, Nachtmann and
Sch\"opf~\cite{Schafer} proposed the use of diffractive
hadron-hadron collisions, where a high-mass central system is
produced, as a way of producing the Higgs boson. Bjorken recognised
such quasi-elastic reactions, in which the protons do not break
despite the appearance of the high-mass system, as a ``superb''
channel to produce exotic particles~\cite{BJ}. The first evaluation
was done in the Higgs case by Bialas and Landshoff~\cite{BL}. It
relied heavily on the use of non-perturbative propagators for the
gluons, which provide an automatic cut-off of the infrared region.
That calculation was then repeated by two of
us~\cite{Cudell:1995ki}, where we showed that by properly treating
the proton form factors, one could use either perturbative or
non-perturbative propagators. However, even with the use of
form factors screening the long wavelengths of the exchanged gluons,
one remained sensitive to the infrared region, as the gluons had a typical
off-shellness of the order of 1 GeV.
Berera
and Collins~\cite{Berera} then calculated the double-pomeron jet
cross sections using a perturbative QCD framework. They identified
large absorptive corrections --- corresponding to the ``gap survival
probability'', {\it i.e.} to multiple pomeron exchanges --- as well
as large virtual corrections --- coming from the large difference of scales
at the jet vertex, and resulting in a ``Sudakov form factor''. However,
they did not model these corrections.

Since then the subject of diffractive dijet and Higgs production has
been extensively discussed from a variety of points of
view~\cite{Kharzeev:2000jwa}--~\cite{Bzdak}. All the calculations
essentially follow the same pattern; the variations come in the
following four ingredients:
\begin{enumerate}
\item \label{ingred1} The two jets are at low rapidity, and at high 
transverse energy $E_T$. They then come
dominantly from gluon production at large transverse momentum (which
we shall note $\bk_2$) and the partonic amplitude can be calculated
in perturbative QCD. Furthermore, the two outgoing gluons are
constrained to be in a colour-singlet state, and, to prevent colour
flow in the $t$ channel, an extra screening gluon is exchanged (Fig.
1.a). The exact relation between the parton $\bk_2$ and the jet $E_T$
will be discussed in Section~\ref{subsec-properties}.
\item \label{ingred2} As the gluons go from low transverse momenta in the 
proton to high ones in the jet system,
there are enhanced double logarithms from vertex corrections (Fig.
1.b), which must be resummed to give the Sudakov form factor.
\item \label{ingred3} The infrared divergences of the gluon propagators 
linked to the proton are regulated
via the use of an impact factor \cite{Cudell:1995ki}, as in Fig.~1.c, or of
non-perturbative propagators \cite{Bzdak}, or via the Sudakov form
factor linked to the hard scattering \cite{Durham}.
\item \label{ingred4} Finally, it is possible that the two protons 
interact at long distances. This leads to
screening corrections, as sketched in Fig. 1.d,
and to a gap survival probability given by the average
squared norm of the $S$-matrix element $|\langle pp|S|pp\rangle |^2$
\cite{Troshin,Frankfurt}.
\end{enumerate}

\begin{figure}
 \centering
 \includegraphics[width=9cm, bb=0 0 522 406]{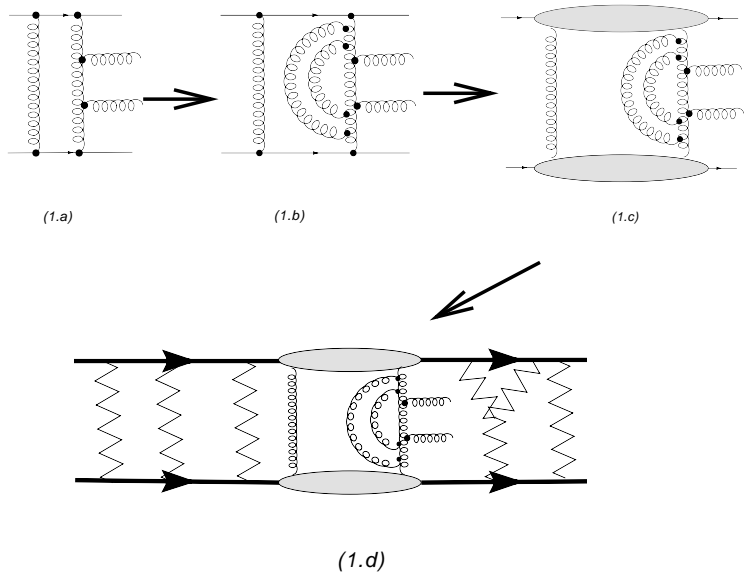}
 \caption{A sketch of the various steps of the calculation.}
 \label{fig:steps}
\end{figure}

In this paper we want to evaluate, in view of the CDF data, the
uncertainties in the various ingredients of the calculation of
central exclusive dijet production. As discussed above, ours is not
the first such calculation, however we believe to shed light on
several important points.  We show that the calculation still lies
mostly in the non-perturbative region, and that many standard
approximations cannot be justified. In particular, we find the
following:
\begin{itemize}
\item At the Born level, exact transverse kinematics is important.
That is to say, the momentum transfers to the hadrons cannot be
neglected with respect to the momentum of the screening gluon, which makes
the exchange colour-neutral. In
Section~\ref{sec-pQCD}, we present a detailed account of the part of
the calculation which is under control, i.e. lowest-order
$qq\rightarrow qqgg$ via a colour-singlet exchange. There we take
the exact transverse kinematics into account.
Furthermore, diagrams in which the screening gluon participates in
the hard sub-process cannot necessarily be neglected, as has been
assumed in previous works. In Section~\ref{subsec-screening}, we show
they could be important once the Sudakov suppression is taken into
account. These issues affect the calculation of the partonic
amplitudes mentioned above in ingredient~\ref{ingred1}.
\item The leading and subleading logs that are resummed to give the
Sudakov form factors are not dominant for the momentum range of the
data, {\it i.e.} the constant terms are numerically important. This
affects the calculation of double logarithm vertex corrections
mentioned above in ingredient~\ref{ingred2}. We discuss the Sudakov
form factor and the problems associated with the large virtual
corrections in Section~\ref{sec-sudakov}.
\item The colour neutrality of the protons has to be implemented
independently of the Sudakov suppression.
This affects the calculations~\cite{Durham} that make use only
of the Sudakov form factor to regulate the infrared divergences of
the gluon propagators mentioned in ingredient~\ref{ingred3}. This is
discussed in Section~\ref{sec-embedding}, where we consider various
ways to embed the perturbative calculation into a proton.
\end{itemize}
In addition to these points, there remains the issue of gap
survival, ingredient~\ref{ingred4} in the list above. Providing an
accurate numerical estimate of this is beyond the scope of this
paper but we do discuss the gap survival probability and current
estimates in Section~\ref{sec-gapsurvival}.

In view of the above uncertainties, we try to outline a few
scenarios that do reproduce the dijet data, and that can be extended
to Higgs production at the LHC. After a few extra corrections, we
give in Section~\ref{sec-roughestimate} a simple estimate of the
cross section, followed in Section~\ref{sec-numerical} by detailed
numerical results. Finally in
Section~\ref{sec-conclusion} we present our conclusions.

\section{The lowest-order perturbative QCD calculation}\label{sec-pQCD}

\subsection{Kinematics}\label{subsec-kinematics}

\begin{figure}
\begin{center}
\includegraphics[height=3cm]{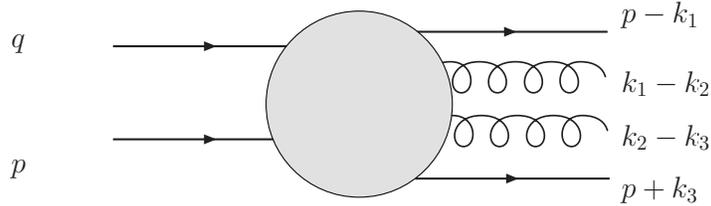}
\caption{Kinematic conventions for the central two-gluon production.}
\label{diag0}
\end{center}
\end{figure}

The backbone of the central quasi-elastic production of two
high-$E_T$ jets is the partonic subprocess $qq\to qqgg$, in which
the produced colour-singlet two-gluon system
 is separated by large rapidity gaps from the two scattering quarks.
The kinematic conventions are shown in Fig.~\ref{diag0}.
We assume that the quarks are massless and consider the collision in
a frame where the incoming quarks have no transverse momenta.
Their momenta $q^\mu$ and $p^\mu$, with $s\equiv 2p\cdot q$,
will be used as the lightcone vectors
for the Sudakov decomposition of all other momenta entering the calculation.
The momentum transfer to the first and second quarks are $-k_1$ and 
$k_3$, respectively,
and are dominated by their transverse parts $-\bk_1$ and $\bk_3$
(we write all transverse vectors in bold).
The momenta of the two produced gluons are
\bea
r_1^\mu = \alpha_1 p^\mu + \beta_1 q^\mu + (\bk_1 - \bk_2)^\mu\,,
\quad \alpha_1 \beta_1 s = (\bk_1 - \bk_2)^2,\nonumber\\
r_2^\mu = \alpha_2 p^\mu + \beta_2 q^\mu + (\bk_2 - \bk_3)^\mu\,,
\quad \alpha_2 \beta_2 s = (\bk_2 - \bk_3)^2\,.\label{ggmomenta}
\eea
The largest contribution to the cross section will come from the region
where longitudinal components obey
\be
1 \gg \beta_1,\, \beta_2 \gg {\bk_i^2 \over s}\,,\quad
1 \gg \alpha_2,\, \alpha_1 \gg {\bk_i^2 \over s}\,,\quad { i=1,\ 2,\ 3}, 
\label{CEPordering}
\ee
as the invariant mass squared of the two-gluon system
\be
M^2_{gg} = (r_1+r_2)^2 = {\left[\beta_2\bk_1 + \beta_1\bk_3 -  
(\beta_1+\beta_2)\bk_2\right]^2\over \beta_1\beta_2}\,
\label{M2gg}
\ee
is much smaller than $s$.
The differential cross section can then be written as a convolution over 
a phase space factorised
between light-cone and transverse degrees of freedom:
\be
d\sigma = {1 \over 16 s^2\, (2\pi)^8}\,{d\beta_1 \over\beta_1}{d\beta_2 
\over\beta_2}\,
d^2\bk_1\,d^2\bk_2\,d^2\bk_3\cdot |{\cal M}|^2\,.\label{dsigma}
\ee
The longitudinal phase space can alternatively be written as
\be
{d\beta_1 \over\beta_1}{d\beta_2 \over\beta_2} = {d\beta \over\beta}{d x 
\over x},\quad
\mathrm{where}\quad  \beta=\sqrt{\beta_1\beta_2}\ \mathrm{and}\ x =  
{\beta_1 \over \beta_2}\,.
\label{PSalt}
\ee
If the two gluons are both integrated in the whole available phase space,
then an extra $1/2$ should be put in the expression of the cross section 
due to Bose statistics.

\subsection{Simplifications for the imaginary part}\label{subsec-imaginary}
As we shall see, the lowest-order calculation will lead to an
amplitude which grows linearly with $s$. As the exchange is $C=+1$,
the amplitude is then mostly imaginary, and  can be calculated via their
standard cuts. At the same lowest order, the real part is suppressed
by a power of $s$, however it will be only logarithmically
suppressed at higher orders. In principle, it can be obtained via
dispersion relations, but we do not concern ourselves with its
contribution, as it will be much smaller than the large uncertainties
in the other parts
of the calculation.

\begin{figure}
\begin{center}
\includegraphics[height=4cm]{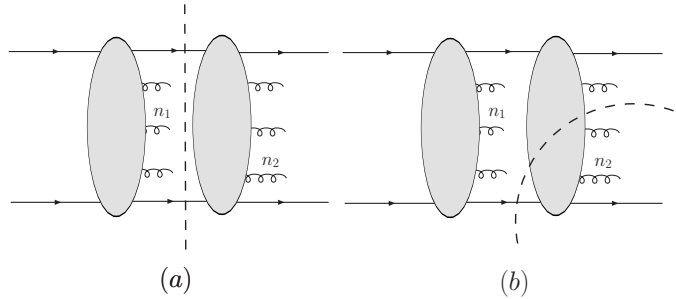}
\caption{Schematic representation of the imaginary part of the production 
amplitude of $n_1+n_2$ gluons
in the central region. $s$-channel cuts of the diagrams, such as (a) 
contribute to the imaginary part, while
``wrong cuts'', such as (b), do not.}
\label{diag1}
\end{center}
\end{figure}

In contrast to central Higgs production, two gluons can be emitted
from all parts of the diagrams in many different ways, which is
represented by the grey circle of Fig.~\ref{diag0}. However, if one
calculates the imaginary part of the amplitude, then there are
multitudinous cancellations among different contributions due to the
positive signature and colour-singlet nature of the exchange, as
well as to the presence of large rapidity gaps. We show in
Fig.~\ref{diag1} two typical cut diagrams that give rise to an
imaginary part for $n_1+n_2$ jet production. The dashed line
represents the kinematic cut of the diagram, i.e. it indicates which
propagators are put on shell in the loop integral. The contributions
of the ``wrong cut'' such as, for example, those shown in
Fig.~\ref{diag1}.b cancel one another. This means that one can write
the amplitude as a sequence of two sub-amplitudes, gauge invariant
on their own: $2 \to 2+n_1 \to (2+n_2)+n_1$, as in
Fig.~\ref{diag1}.a. Since in our case $n_1+n_2=2$, there are three
generic situations: ($n_1=0,\, n_2=2$), ($n_1=2,\, n_2=0$), or
($n_1=1,\, n_2=1$).

Each of the sub-amplitudes in Fig.~\ref{diag1}.a describes the
emission of one or two gluons in the central region. In principle,
all the vertices needed for the calculation of two-gluon production
in arbitrary kinematics can be found in the literature. Production
of one gluon can be described by the standard Lipatov vertex
\cite{BFKL}, while emission of two gluons involves an effective
four-gluon vertex in the quasi-multi-Regge kinematics~\cite{RRGG}.
Such non-local vertices take into account gluon emission not only
from the $t$-channel gluons themselves, but also from the quarks. In
the case of large transverse momentum of the produced gluons, which
is the focus of the present paper, the situation simplifies, since
emission from $t$-channel gluons is dominant, and the amplitude is
more conveniently calculated using Feynman diagrams and cutting
rules.

\subsection{Central production of two gluons with large transverse
momentum}\label{subsec-centralgg} 
\begin{figure}
\begin{center}\mbox
{{\includegraphics[width=15cm]{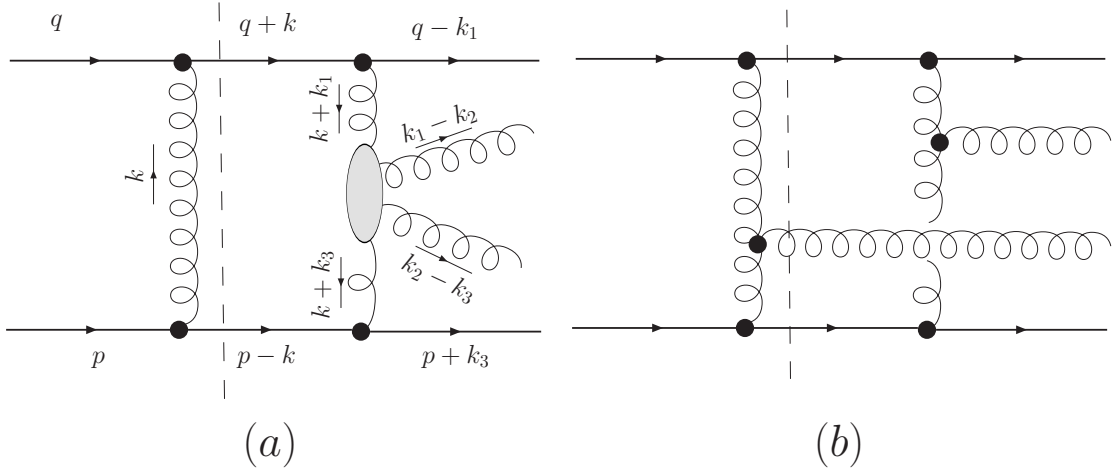}}}
\caption{(a) A generic diagram for $qq\to q+gg+q$ scattering in the 
high-$\bk_2^2$ regime; the shaded
blob represents the diagrams for the elastic $gg$ scattering. (b) A diagram
with two gluons emitted from two $t$-channel legs is suppressed by an 
extra $1/\bk_2^2$ factor. }
\label{diag2}
\end{center}
\end{figure}

We are interested in the
quasi-elastic production of two gluons with large relative
transverse momenta of the order of tens of GeV. The requirement that the
protons remain intact effectively cuts the differential cross
section at small values of momentum transfers, $\bk_1^2, \, \bk_3^2
\lsim 1/B_p$, where $B_p$ is a typical proton elastic slope in
hadronic reactions (for more discussion on what numerical value for
$B_p$ would be most appropriate in our case, see
Section~\ref{subsec-embedding-ugd}). Therefore, $\bk_1^2, \, \bk_3^2
\ll \bk_2^2$, and the process can be viewed as a collision of two
nearly-collinear but energetic gluons\footnote{Strictly speaking, in
the lab frame, one of these two gluons can be very soft and emitted
at large angle relative to the quark collision axis. However, after
an appropriate longitudinal boost the above description will become
true.} $g^*g^*\to gg$ accompanied by an additional exchange of an
extra screening gluon to restore the neutrality of the $t$-channel
colour exchange. The set of diagrams to be considered is then
reduced to those of Fig.~\ref{diag2}, and
to their counterparts where each gluon
is emitted from the other side of the cut.

Lowest-order diagrams with gluons emitted from different $t$-channel
gluons, such as the one of Fig.~\ref{diag2}.b, are
suppressed\footnote{Note that the presence of hard transverse
momentum in the $s$-channel partons in these diagrams changes
nothing since these partons are on-shell, so that the transverse
momentum does not suppress the amplitude.} by one extra power of
$\bk_2^2$. The situation may become different at higher orders, as
we shall explain in Section~\ref{subsec-screening}. Also, as we are
calculating cut diagrams, the sub-amplitudes are at the tree level
and there are no ghost contributions.

The Mandelstam invariants in the two-gluon collision can be written
\be
M^2_{gg}\equiv s_{gg} \approx \bk_2^2 {(\beta_1 + \beta_2)^2 \over 
\beta_1 \beta_2}\,,\quad
t_{gg} \approx -\bk_2^2 {\beta_1 + \beta_2 \over \beta_1}\,,\quad
u_{gg} \approx -\bk_2^2 {\beta_1 + \beta_2 \over \beta_2}\,.
\ee
Note that these invariants depend only on the ratio
$x=\beta_1/\beta_2$, which is related to the difference of the
rapidities of two produced gluons.

Let us also introduce the momenta of the colliding gluons
\be
\kappa_1^\mu = (k+k_1)^\mu\,,\quad \kappa_3^\mu = -(k+k_3)^\mu\,,\label{kappa4}
\ee
The imaginary part of the amplitude can be represented as
\be
\mathrm{Im}{\cal M} = {g^4 \over 4\pi^2}{\delta^{ab}\over 4N^2}\int {d^2 
\bk \over \bk^2
 \bkappa_1^2 \bkappa_3^2}\cdot
\sum_{\lambda_i}j^{(1)*}_{\lambda_1}j^{(2)*}_{\lambda_2}
{M}_{\lambda_1\lambda_2\to \lambda_3\lambda_4}\,.\label{ampEPA}
\ee
where
\be
j^{(1)*}_{\lambda_1} = \bar u(q+k) \gamma\cdot e^{*}_{\lambda_1} u(q-k_1) 
\,,\quad
j^{(2)*}_{\lambda_2} = \bar u(p-k) \gamma\cdot e^{*}_{\lambda_2} u(p+k_3)
\ee
are the amplitudes $q\rightarrow qg$ with gluon polarisation vectors 
discussed below.
Strictly speaking, the two colliding gluons are virtual. In our 
kinematics, their off-shellnesses are $-\bkappa_1^2$ and
$-\bkappa_3^2$, which are much smaller than $\bk_2^2$. Since the hard 
scale of the scattering sub-process is given by
$\bk_2^2$, one can neglect the non-zero virtualities and approximate the 
amplitude
by $gg\to gg$ scattering of on-shell transversely polarised gluons, and 
neglect the contribution from the
longitudinal polarisations.
The calculation is then manifestly gauge invariant.

The helicity amplitudes for the tree-level scattering of two gluons in a 
colour-singlet state
are most easily calculated in the centre-of-mass frame:
\be
{M}_{\lambda_1\lambda_2\to \lambda_3\lambda_4} = 2 g^2 {N\over N^2-1} 
\delta^{ab}\delta^{c_1c_2}
\left({s_{gg} \over t_{gg}} + {s_{gg}\over u_{gg}}\right)  
e^{i(\lambda_1-\lambda_2)\phi}
A(\lambda_1\lambda_2\to \lambda_3\lambda_4)\,,
\label{Mgg}
\ee
where $\phi$ is the azimuthal angle of the two-gluon-production plane
with respect to the quantisation axis. Note that this quantisation axis is
arbitrary, and changing it will produce changes in $M$ which will be 
compensated by
opposite changes in $j_{\lambda_i}^{(l)*}$.

The fact that Eq.~(\ref{Mgg}) is written in a frame which is different 
from the laboratory frame
does not pose any problem.
Indeed, in order to pass from the laboratory frame to the $gg$ 
centre-of-mass frame
with gluons colliding along the $z$ axis,
one first has to perform a longitudinal boost to make the energies of the 
colliding gluons equal,
then a transverse boost to make the total momentum of two gluons zero,
and then rotate the frame to align the $z$ axis with the direction of the 
incoming gluons.
The large longitudinal boost does not change $\phi$, while the transverse boost
and the rotation by a small angle have negligible effect on the hard 
momentum $k_2$.
Therefore, one can safely understand $\phi$ in Eq.~(\ref{Mgg}) as the 
azimuthal angle in the lab frame.

The non-zero $A(\lambda_1\lambda_2\to \lambda_3\lambda_4)$ in 
Eq.~(\ref{Mgg}) are
\bea
&&A(++\to ++) = A(--\to --) = 1\,,\nonumber\\
&&A(+-\to +-) = A(-+\to -+) = {u_{gg}^2 \over s_{gg}^2}\,,\nonumber\\
&&A(+-\to -+) = A(-+\to +-) = {t_{gg}^2 \over s_{gg}^2}\,.\nonumber
\eea
This list exhibits the total helicity conservation rule, which is
a consequence of the helicity properties of a general tree-level
$n$-gluon scattering amplitudes, see e.g.~\cite{parketaylor}.
In our case it implies, in particular, that $++$ and $--$ amplitudes do not interfere
with any other.

The fact that the colliding gluons are soft, and the momentum hierarchy 
of Eq.~(\ref{CEPordering}),
simplifies the calculation.
Each of the polarisation vectors for the initial gluons can be chosen 
orthogonal to
both $\kappa_1^\mu$ and $\kappa_3^\mu$, and within our accuracy
can be generically written as
\be
e^\mu_{\lambda} = \bbe^\mu_\lambda + {2 \over s}\left[
p^\mu{\bbe_\lambda\cdot\bkappa_1 \over\beta_1+\beta_2}
+q^\mu{\bbe_\lambda\cdot\bkappa_3 \over \alpha_1+\alpha_2}
\right]\,,
\ee
Here, $\bbe_\lambda$ is the standard polarisation vector in the 
transverse plane,
$$
\bbe_\lambda = -{1 \over\sqrt{2}}(\lambda,\, i)\,,
$$
with  $\lambda=\lambda_1$ for the first gluon and
$\lambda=-\lambda_2$ for the second one,
since they move in opposite longitudinal directions.
One can now simplify
\bea
j^{(1)*}_{\lambda_1} &\approx& 2{\bbe_{\lambda_1}^*\cdot\bkappa_1 \over 
\beta_1 + \beta_2}
= -{2 \over \beta_1 + \beta_2}\, {\lambda_1 \over \sqrt{2}} |\bkappa_1|
e^{-i\lambda_1 \phi_1}\,,\nonumber\\
j^{(2)*}_{\lambda_2} &\approx& 2{\bbe_{-\lambda_2}^*\cdot\bkappa_3 \over 
\alpha_1 + \alpha_2}
= -{2 \over \alpha_1 + \alpha_2}\, {-\lambda_2 \over \sqrt{2}} |\bkappa_3|
e^{i\lambda_2 \phi_3}\,,
\eea
where $\phi_1$ and $\phi_3$ are the azimuthal angles of $\bkappa_1$ and 
$\bkappa_3$,
respectively.

Squaring the amplitude, one finds the following expression
\be
\sum_{f} \sum_{\lambda_i,\lambda_i^\prime} j^{(1)*}_{\lambda_1} 
j^{(2)*}_{\lambda_2}
j^{(1^\prime)}_{\lambda_1^\prime}j^{(2^\prime)}_{\lambda_2^\prime}
\, M_{\lambda_1\lambda_2\to f}M^*_{\lambda_1^\prime\lambda_2^\prime\to f}\,,
\label{jjjj}
\ee
where $f$ labels the polarisation states of the final two-gluon system.
Summation over final fermions and averaging over initial ones is also 
implied here.
Note that, in contrast to the standard Weizs\"acker-Williams approximation,
where the initial particle momentum is the same in $j^*_\lambda$
and $j_{\lambda^\prime}$, here it is different due to $\bf k\not = \bf 
k^\prime$.
This induces correlations between the colliding gluons,
and will lead in a moment to the conclusion that the fully unpolarised
$qq \to qq+gg$ cross section  integrated over all phase space
is {\em not} proportional to the unpolarised $gg \to gg$ cross section.

The only non-trivial interference here is between $M_{+- \to f}$ and 
$M_{-+ \to f}$,
with $f = +-$ or $-+$.
Such a term introduces an  azimuthal dependence for high-$|\bk_2|$ gluons 
via the factor
$\exp(4i\phi)$. It contributes to the azimuthal correlations between
the high-$E_T$ jets and the proton scattering planes, but if integrated 
over $\phi$
(still keeping  the cross section differential in $\bk_2^2$), this term 
vanishes.
This allows us to consider only the diagonal contributions in Eq.~(\ref{jjjj}),
$\lambda_1^\prime = \lambda_1$, $\lambda_2^\prime = \lambda_2$.
The result is
\bea
&&\sum_{f} \sum_{\lambda_i} j^{(1)*}_{\lambda_1}j^{(2)*}_{\lambda_2}
 j^{(1^\prime)}_{\lambda_1}j^{(2^\prime)}_{\lambda_2}
\, |M_{\lambda_1\lambda_2\to f}|^2 \nonumber\\
&& =\ 8{|\bkappa_1||\bkappa_3| |\bkappa_1^\prime||\bkappa_3^\prime|
\over (\beta_1+\beta_2)^2 (\alpha_1+\alpha_2)^2}
\left[
|M_0|^2\cos(\phi_1 - \phi_3 - \phi_1^\prime + \phi_3^\prime)
+ |M_2|^2\cos(\phi_1 + \phi_3 - \phi_1^\prime - \phi_3^\prime)
\right]
\nonumber\\
&& =\ 8{|\bkappa_1||\bkappa_3| |\bkappa_1^\prime||\bkappa_3^\prime|
\over (\beta_1+\beta_2)^2 (\alpha_1+\alpha_2)^2}
\biggl[
(|M_0|^2 + |M_2|^2)\cos(\phi_1 - \phi_1^\prime)\cos(\phi_3 -   
\phi_3^\prime)\label{jjjj2}\\
&&\qquad \qquad
+\  (|M_0|^2 - |M_2|^2)\sin(\phi_1 - \phi_1^\prime)\sin(\phi_3 - \phi_3^\prime)
\biggr], \nonumber\\
&&
=\ { 8 \over (\beta_1+\beta_2)^2 (\alpha_1+\alpha_2)^2}
\biggl\{(|M_0|^2 + |M_2|^2)(\bkappa_1 \cdot \bkappa_1^\prime) 
(\bkappa_3\cdot  \bkappa_3^\prime)\nonumber\\
&& \qquad \qquad +\ (|M_0|^2 - |M_2|^2)[(\bkappa_1\cdot \bkappa_3) 
(\bkappa_1^\prime\cdot \bkappa_3^\prime) -
(\bkappa_1\cdot \bkappa_3^\prime)(\bkappa_1^\prime\cdot \bkappa_3)]\biggr\}.
 \label{jjjj3}
\eea
Here, $|M_0|^2$ and $|M_2|^2$ are the amplitudes squared of elastic collision of
two gluons with total helicity $\lambda=\lambda_1-\lambda_2=0$ or 2 summed over the final helicity states:
\bea
|M_0|^2&\equiv& {1\over 2}\left[|M_{++\to ++}|^2 +|M_{--\to 
--}|^2\right],\nonumber\\
|M_2|^2&\equiv& {1\over 2}\left[|M_{+-\to +-}|^2 +|M_{+-\to 
-+}|^2+|M_{-+\to -+}|^2+|M_{-+\to +-}|^2\right].
\eea
Note that if $\bk$ were equal to $\bk^\prime$,  then $\phi_i$ would be 
equal to $\phi_i^\prime$,
and one would end up with the unpolarised cross section, $\propto |M_0|^2 
+ |M_2|^2$ in Eq.~(\ref{jjjj2}),
in the spirit of the usual Weizs\"acker-Williams approximation.
Another observation is that in the limit
$$
|\bk_1|,\, |\bk_3| \ll |\bk|,\, |\bk^\prime|\,,
$$
the angles are $\phi_3 \approx \phi_1 + \pi$ and $\phi_3^\prime \approx 
\phi_1^\prime + \pi$,
so that Eq.~(\ref{jjjj2}) would simplify to
\be
8{\bk^2 \bk^{\prime 2}
\over (\beta_1+\beta_2)^2 (\alpha_1+\alpha_2)^2}
\left[
|M_0|^2 + |M_2|^2\cos(2\phi_{\bk} - 2\phi_{\bk'})
\right]\,.
\ee
After the azimuthal averaging over $\phi_{\bk}$ and $\phi_{\bk'}$, the 
$|M_2|^2$ term
vanishes and only the amplitude with $\lambda=0$  contributes,
so that in this limit one obtains the $J_z=0$ rule \cite{Durham}.

Finally, since
$$
{1 \over (\beta_1+\beta_2)(\alpha_1+\alpha_2)}\left({s_{gg} \over t_{gg}}  
+ {s_{gg}\over u_{gg}}\right)
= {s \over \bk_2^2}\,,\nonumber\\[2mm]
$$
one can write the partonic differential cross section as
\bea
d\sigma_{qq} & = & {1 \over 9\pi^6}\left({g^2 \over 4\pi}\right)^6  
{d\beta_1 \over \beta_1} {d\beta_2 \over \beta_2}
{d^2\bk_1 d^2\bk_2 d^2 \bk_3 \over (\bk_2^2)^2}
\int { d^2\bk \over \bk^2 \bkappa_1^2 \bkappa_3^2} { d^2\bk^\prime \over 
\bk^{\prime 2} \bkappa_1^{\prime 2} \bkappa_3^{\prime 2}}
\ \bkappa_1^\mu \bkappa_3^\nu \bkappa_1^{\prime\tau}  
\bkappa_3^{\prime\sigma}\label{partonic-xsection}\\
&\times&\Biggl\{
\biggl[g_{\mu\tau}g_{\nu\sigma}+g_{\mu\nu}g_{\tau\sigma}  
-g_{\mu\sigma}g_{\nu\tau}\biggr]
+
\biggl[g_{\mu\tau}g_{\nu\sigma}-g_{\mu\nu}g_{\tau\sigma}  
+g_{\mu\sigma}g_{\nu\tau}\biggr]
\left({u^4_{gg} \over s^4_{gg}} + {t^4_{gg}\over s^4_{gg}}\right)\Biggr\}
\,,\nonumber
\eea
where the first term in brackets corresponds to $\lambda=0$, and the 
second one to $\lambda=2$.

\section{The Sudakov form factor}\label{sec-sudakov}

\subsection{Resummation}\label{subsec-sudakov-resum}
\begin{figure}
\begin{center}
{\includegraphics[width=10cm]{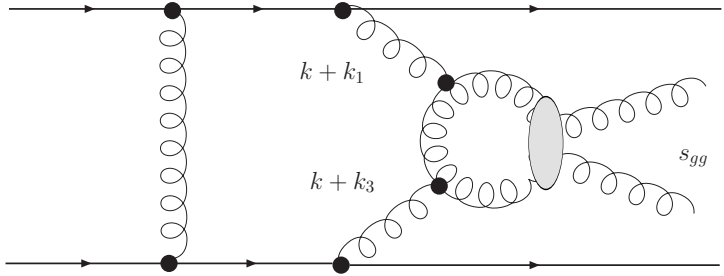}}
\caption{A higher-order diagram for $qq\to q+gg+q$ scattering leading to 
large doubly logarithmic
corrections in the Feynman gauge. }
\label{sud}
\end{center}
\end{figure}

In the diagrams of Fig.~\ref{diag2}.a, the gauge field goes from a 
long-distance
configuration to a short-distance one. This is the situation for which large
doubly logarithmic corrections, \hbox{$\sim\log^2(\bk_2^2)$}, are 
expected, from virtual diagrams such as
those of Fig.~\ref{sud}. These corrections are there for any gauge theory 
\cite{sudabooks}
and have been calculated in QCD in \cite{Dokshitzer:1978hw}. For initial 
gluons on-shell,
these corrections actually diverge, and this divergence is cancelled by 
the  bremsstrahlung diagrams
in inclusive cross sections. Hence, the cancellation of infrared 
divergences
means that the logarithmic structure of the virtual corrections is 
identical to that of the
bremsstrahlung diagrams\footnote{ For the case at hand, the infrared 
region is cut off by the off-shellness of the
initial gluon, so that the logarithms are large, finite and of the order of
$ \log(s_{gg}/ (\bk_1+\bk)^2)\log(s_{gg}/ (\bk_3+\bk)^2)\geq 25$ for the 
CDF exclusive jet production.}.

So there are two interpretations of the Sudakov form factor.
On the one hand, one can view it as a resummation of the double-log  
enhanced virtual corrections.
For example, at the one-loop order, these corrections include diagrams  
with the integrals
over the fraction of the light-cone momentum, $z$, and the transverse 
momentum, $\bq^2$,
of the  particle in the loop. Each of these integrals builds up a 
logarithm coming
from regions $\Delta \ll z$ and $\ell^2 \ll \bq^2 \ll \mu^2$, respectively.
Here, $\Delta$ is a cut-off to be discussed below, $\ell^2$ is a typical 
virtuality
of the initial gluons in the $g^*g^*\to gg$ process, and $\mu$ is the 
scale of the hard process,
which is of the order of the jet transverse energy $E_T$.

On the other hand, the cancellation of logarithms between virtual and 
real diagrams means that
same Sudakov form factor can be reworded in a Monte-Carlo
language as the probability of not emitting any extra partons.
In this interpretation, one considers collision of two gluons and calculates
the probabilities that a hard sub-process is accompanied by a certain number
of secondary partons. Subtracting from unity all the probabilities of emission
gives the probability of the purely exclusive reaction.

Using this technique, one can then use for the Sudakov form factor the 
expression coming from
Monte Carlo simulations \cite{MCbook}:
\be
T (\mu^2,\ell^2)= \exp\left[-S(\mu^2,\ell^2)\right]\,,\quad
S(\mu^2,\ell^2)=\int_{\ell^2}^{\mu^2} {d\bq^2 \over \bq^2}  
{\alpha_s(\bq^2) \over 2\pi}
\int_0^{1-\Delta} dz \left[zP_{gg}+N_f P_{gq}\right]\,.\label{sudasplitting}
\ee
Here,  the lower scale $\ell^2$ is understood as the virtuality from 
which the evolution starts, and
$P_{gg}$ and $P_{gq}$ are the unregularised splitting functions,
\be
P_{gg}(z) = 2 N_c \left[{z \over 1-z} + {1-z \over z} + z(1-z)\right]\,,\quad
P_{gq}(z) = {1 \over 2}\left[z^2 + (1-z)^2\right]\,.
\ee
We shall refer to this form of the vertex corrections as the Splitting 
Function Approximation~(SFA).

If instead of  $\alpha_s(\bq^2)$, we take $\alpha_s(\mu^2)$ in 
(\ref{sudasplitting}), we can easily work out the
double logarithm approximation (DLA), which comes from the $2N_c/(1-z)$ 
term in $P_{gg}$:
\be
S_{DLA}(\mu^2,\ell^2) = {3 \alpha_s \over \pi} \int_{\ell^2}^{\mu^2}
{d\bq^2 \over \bq^2} \log\left({1\over \Delta}\right)\,.
\ee
The coefficient in front of the double logarithm depends on the 
definition of $\Delta$.
First, pure DGLAP kinematics leads to the cut-off $\Delta = \bq^2/\mu^2$.
Using this cut-off, one obtains
 the expression worked out in \cite{Dokshitzer:1978hw}.
However, it can be argued that the coherent effects lead to angular ordering
in the successive splitting of the secondary partons \cite{CCFM}. This 
ordering introduces
a more restrictive limit on the $z$ integral, which becomes linear in 
$|\bq|$: $\Delta = |\bq|/\mu$.
With this definition of $\Delta$, one obtains a double-log result,
which is twice smaller than in \cite{Dokshitzer:1978hw}:
\be
S_{DLA}(\mu^2,\ell^2) = {3 \alpha_s \over 4\pi} \log^2\left({\mu^2\over 
\ell^2}\right)\,.\label{sudakDLL1}
\ee
Numerically, this change is very substantial; it can easily lead to 
factors ${\cal O}(10)$
in the cross section.
In the following, we shall only use the latter prescription for $\Delta$.
Note that, together with formula (\ref{sudasplitting}),
it can also be derived from the CCFM equation \cite{CCFM2}.

\begin{figure}
\begin{center}
{\includegraphics[width=10cm]{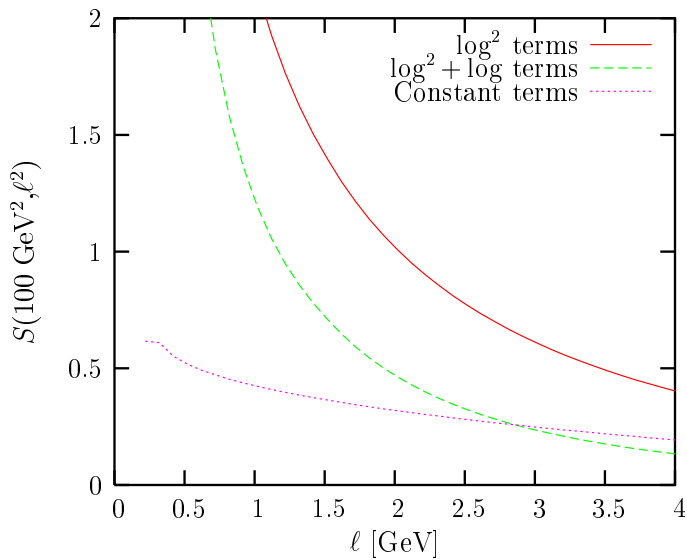}}
\caption{The contribution of the various terms entering the argument of 
the Sudakov form factor. }
\label{fig-sud2}
\end{center}
\end{figure}

Several comments are in order. First of all, Eq.~(\ref{sudasplitting}) 
does not include all the
single logarithms present in the process, which, to the best of our 
knowledge, have been evaluated only for the
quark vertex \cite{sudalogs}. One does not know in the present case whether
further single-logarithmic corrections are large, or whether they break 
the exponentiation.
Moreover,even if one takes Eq.~(\ref{sudasplitting}) at face value,
its validity depends on the fact that the logarithms are dominant.
We show in Fig.~\ref{fig-sud2} the contribution to $S$ of the constant 
terms, compared to
that of  logarithms, for $\mu=10$ GeV. For typical scales relevant for 
the CDF measurements
($\mu \sim {\cal O}$(10 GeV), $\ell \approx 1$ GeV) the single-log corrections
are large. They are negative for $P_{gg}$ and positive for $P_{qg}$.
One can even say that the overall contribution of $g\to q\bar q$ splitting
is equally important as $g \to gg$. As we explained above, these 
logarithmic corrections
are only an educated guess, and the fact that they are huge makes the 
theoretical predictions
unstable.

Furthermore, as can also be seen from Fig.~\ref{fig-sud2}, it turns out 
that even the constant terms
in (\ref{sudasplitting}) are numerically important. One sees that, for an 
upper scale of 10 GeV,
the logarithms are dominant only
for $\ell\ll 2.5$ GeV, which means that we can trust the perturbative 
formula~(\ref{sudasplitting})
only in the non-perturbative region!

Our  analysis implies that any other source of single-log
contributions (for example, the one discussed in
Section~\ref{subsec-screening} below, or a mere shift of the upper
and lower scales in the logarithm) is also expected to affect
strongly the resulting value of the Sudakov form factor. In
Section~\ref{subsec-uncertainties} we will study this sensitivity in
detail.

\subsection{What is the upper scale in the Sudakov 
integral?}\label{subsec-upperscale}

The standard discussion of the Sudakov form factors relies almost exclusively
on corrections to the quark electromagnetic form factor.
In this case, the hard vertex is characterised by
the single kinematical quantity $s_{q\bar{q}}$.
In the Monte Carlo language, this energy defines the phase space available
for the secondary parton emission which needs to be suppressed.
This is why the upper scale in the Sudakov integral can be taken as
$\mu^2 = c\cdot s_{q\bar q}$, with numerical coefficient $c = {\cal O}(1)$.

The same argument applies to the Higgs central exclusive production.
In this case, the hard vertex is effectively point-like,
as the transverse momenta and virtualities of the top quarks
inside the loop are larger or of the order of the Higgs mass.

With dijets with large invariant mass, we enter a new kinematical regime.
The two gluons can have large invariant mass $s_{gg}$ both via large 
transverse momentum
or via strong longitudinal ordering, see (\ref{M2gg}).
In the former case, the transverse momentum exchange in the $gg \to gg$ 
scattering is large,
$\bk_2^2 \sim s_{gg}$, while in the latter case, $\bk_2^2 \ll s_{gg}$.

A typical double-log-enhanced correction to some vertex $V$ gets one 
logarithm from the fraction
of the lightcone variable and another one from the transverse momentum or 
virtuality integral.
In order for the transverse momentum integral to produce a sizable log,
there should exist a large transverse momentum or large virtuality inside
the effective vertex. In this case the transverse momentum in the loop, 
when flowing through $V$
does not change its amplitude. In the Monte Carlo language,
in order for the backwards evolution to develop, one must have a large 
initial hard virtuality inside the vertex.
On the contrary, if the vertex does not involve any
large transverse momentum or large virtuality, then the transverse loop 
integral
is suppressed by the vertex $V$, and the transverse logarithm does not build up.
Once the structure of the vertex $V$ is resolved by the incoming gluons, 
the structure
of the amplitude changes, and double logarithms disappear. Because this 
transition is
not sharp, the actual value of the upper scale is somewhat uncertain, but 
it must be
of the order of $\bk_2^2$.

This can be immediately seen in the extreme case of production of two gluons
in multi-Regge kinematics (and therefore large $s_{gg}$) but with small 
transverse momenta,
of the order of typical loop transverse momentum in the BFKL ladder.
The amplitude of the $gg\to gg$ subprocess is then $\propto s_{gg}/t_{gg}$,
and this process does not involve any hard gluon.
Consider now a loop correction to it.
If the loop integral involves large transverse momentum $\bq$, then it 
will flow through the ``original''
$gg \to gg$ subprocess and will suppress its amplitude to $\propto 
s_{gg}/\bq^2$.
This suppression prevents the development of the transverse momentum logarithm,
in accordance with the general BFKL theory.
It is in this suppression of the vertex where the key difference lies
between the usual Sudakov correction to the quark electromagnetic
form factor with its point-like vertex (or the Higgs production) and the 
dijet production.

With this discussion in mind, we believe that the physically motivated 
choice for the upper scale in the Sudakov integral
should be related with the transverse momentum transfer in the $gg\to gg$ 
subprocess, but not with $s_{gg}$.

As for the present study, with dijets produced in the central rapidity 
region and not in multi-Regge kinematics,
these two prescriptions do not lead to any major difference in the 
$E_{T,\,min}$ distribution of the cross section.
However, they will lead to vastly different $M^2_{jj}$ shapes of the 
cross section, which will be discussed
in Section~\ref{subsec-properties}.

\subsection{The role of the screening gluon}\label{subsec-screening}

It is usually assumed that the extra gluon in the $t$-channel is needed only
to screen the colour exchange and does not participate in the hard 
sub-process, Fig.~\ref{diag2}.a.
So the pomeron fusion is very similar to gluon fusion (up to colour 
factors), and the details
of the final state of the hard interaction do not substantially change 
the calculation.
This would make dijet exclusive production essentially identical to Higgs 
exclusive production,
and leads to the hope that the CDF data can be used
to calibrate Higgs production.
At the lowest order, this assumption is justified,
since the emission of two gluons from two different $t$-channel legs, 
Fig.~\ref{diag2}.b,
leads to an extra $1/\bk_2^2$ suppression.

The virtual corrections discussed so far in connection with the Sudakov 
form factor do not lead  to any non-trivial
``cross-talk'' between the two $t$-channel gluons. Besides, the standard 
BFKL-type exchanges can
be re-absorbed in the evolution of the gluon density, as we shall see in 
the next section,
so that the screening gluon effectively decouples from the hard sub-process.

Here, we would like to discuss two potential mechanisms through which the 
screening gluon
may get involved in the dynamics.

The first one is specific to dijet production.
We know that the Born-level amplitude of the standard diagram 
Fig.~\ref{diag2}.a must be corrected by a
Sudakov form factor. Keeping only the leading powers of the transverse 
momenta, we can write
\be
\mathrm{Im}{\cal M}\sim {1\over \bk_2^2} \int {d^2\bk \over  (\bk^2)^3} 
\bk^2 \exp\left[-S(\bk_2^2,\bk^2)\right] \sim
{1\over \bk_2^2} {1 \over \langle\bk^2\rangle}  
\exp\left[-S(\bk_2^2,\langle\bk^2\rangle)\right]\,,\label{ampestimate1}
\ee
where $\langle\bk^2\rangle \approx {\cal O}(1\mbox{ GeV}^2)$ is the 
position of the saddle point.

\begin{figure}
\begin{center}
\includegraphics[height=3cm]{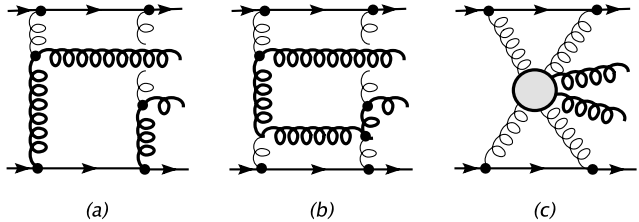}
\caption{At the Born level, (a), the diagram with two gluons emitted from 
different legs
involves hard momentum flow in the $t$-channel, represented by thick 
gluon lines. As a result,
the one-loop corrections, (b), are never double-log enhanced. However, 
they ``localise''
the hard momentum flow in the diagram, and successive loop corrections
are now double-log enhanced. Their calculation amounts to calculation of
a novel type of Sudakov form factor to the irreducible $gggg\to gg$ 
process, (c).}
\label{fig-newsudak}
\end{center}
\end{figure}

Let us now consider similar corrections to the suppressed diagram of 
Fig.~\ref{diag2}.b,
the hard part of which is shown in Fig.~\ref{fig-newsudak}.a.
At the one-loop level, corrections to these diagrams are not double-log 
enhanced\footnote{We
remind the reader of the QED result for the Sudakov form factor in $q\bar 
q$ collision
with quark virtualities $k_1^2$ and $k_2^2$ and total momentum $q$, 
\cite{sudabooks,QEDSudakov-offshell}:
$$
S = {\alpha \over 2\pi} \log\left|{q^2\over k_1^2}\right| 
\log\left|{q^2\over k_2^2}\right|\,.
$$
In a very asymmetric case, with $p_1^2 \ll q^2$ and $p_2^2 \sim q^2$,
this expression is only single-log enhanced.}.
One can however channel the hard transverse momentum flow as shown in
Fig.~\ref{fig-newsudak}.b. This $\alpha_S$ correction is enhanced by a 
single log.
After that, however, the lines corresponding to the hard process remain 
the same at higher orders, and the outer gluon legs
do not carry any hard momentum, see Fig.~\ref{fig-newsudak}.c.
Thus, successive loops do produce double-log corrections to the 
irreducible $gggg\to gg$ vertex,
and, perhaps, may even be resummed, leading to a novel type of Sudakov 
form factor,
the calculation of which might be interesting on its own.
However, we stress that the corrections to the lowest-order result always 
lack one logarithm,
as they are of order $\alpha_S \log(\bk_2^2) \, (\alpha_S \log^2(\bk_2^2))^n$.

Arguably, this is an indication that the corresponding factor that 
accompanies the Born-level amplitude,
which we write as $\exp[-S_{new}(\bk_2^2,\bk^2)]$, is not as small as the 
Sudakov form factor: it might be that
 $S_{new}(\bk_2^2,\bk^2) \ll S(\bk_2^2,\bk^2)$.
A very rough estimate of the resulting amplitude is (see derivation in 
Appendix A)
\be
\mathrm{Im}{\cal M}_{new}\sim {1\over (\bk_2^2)^2} \int {d^2\bk \over \bk^2} \,
\exp\left[-S_{new}(\bk_2^2,\bk^2)\right] \,.\label{ampestimate2}
\ee
Note that in contrast to the ``standard'' amplitude (\ref{ampestimate1}), 
the integrand here
extends to much higher values of $\bk^2$.
Therefore, it might happen that, after all, the diagram 
Fig.~\ref{diag2}.b is not as much suppressed as it looks at the 
Born-level. Certainly, for a very hard process, {\it 
i.e.} for $\bk_2^2\rightarrow\infty$,
it can be safely neglected.
However, its importance grows at smaller $E_T$, and
it is not obvious without a detailed calculation  from what values of 
$E_T$ the estimates from Eq.~(\ref{ampestimate1})
and Eq.~(\ref{ampestimate2}) become of the same order, and whether this 
interval includes
the CDF kinematic region.
As the $t$-channel gluons are in a colour-singlet
state,  the diagrams of Figs.~\ref{diag2}a and b have opposite signs, so 
that the overall effect
will be to decrease the jet cross section w.r.t. the Higgs or 
$\gamma\gamma$ production cross sections.

\begin{figure}
\begin{center}
\epsfig{file=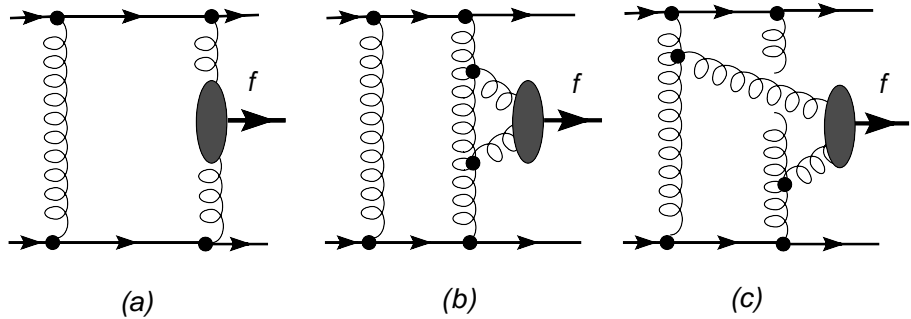,width=12cm}
\end{center}
\caption{A colourless final state $f$ can be produced not only via the 
standard diagram (a) but also via
a collision of two gluons produced in multi-Regge-kinematics, (b) and (c).
The diagram (b) is the first term in the Sudakov form factor,
while diagram (c) is specific for central exclusive production
 and modifies this form factor in the subleading logarithms. Note that 
the hard process is
totally concentrated in the blob and that no gluon line
carries large transverse momentum.}
\label{fig-secondcorrection}
\end{figure}

The above corrections are specific to the dijets and are absent
for Higgs or $\gamma\gamma$~production.
Let us now discuss another potential
correction that affects all of these final states (which we denote 
generically $f$)
and, even more importantly, which is not suppressed by a power of the 
hard scale as was (\ref{ampestimate2}).

If a process $gg \to f$ is possible, then one can produce $f$ not only 
via the standard
mechanism, Fig.~\ref{fig-secondcorrection}.a, but also via processes
Fig.~\ref{fig-secondcorrection}.b and Fig.~\ref{fig-secondcorrection}.c.
Note that none of the gluons shown in these diagrams carries hard   
transverse momentum
of order $E_T$. Diagrams Fig.~\ref{fig-secondcorrection}.b and 
Fig.~\ref{fig-secondcorrection}.c
possess an extra $\alpha_s$ but are enhanced by at least one large logarithm.
More specifically, diagram Fig.~\ref{fig-secondcorrection}.b is 
double-log enhanced and is effectively
taken into account by the Sudakov form factor.
However, the standard Sudakov form factor (the one used in inclusive production
of a colourless state) does not include diagrams like
Fig.~\ref{fig-secondcorrection}.c, which are specific for the central exclusive production.

We stress that the contribution of this diagram is not small, as it is 
single-log enhanced (see Appendix B).
Therefore, such diagrams must be resummed, and they will
lead to modifications to the Sudakov form factor at the single-log level,
that were absent in the inclusive case.
Recalling the very substantial role of single-log effects in the Sudakov 
form factor,
one might expect numerically very sizable corrections due to the diagrams 
like Fig.~\ref{fig-secondcorrection}.c.
Further analysis is definitely needed to bring these corrections under control.

\section{Embedding the gluon production into proton-proton
collisions}\label{sec-embedding}

\subsection{Impact factors}
So far, we have considered colour-singlet quark-quark scattering,
and expression (\ref{partonic-xsection}) is singular when  the
exchanged gluons go on-shell, i.e. $ \bk^2 \bkappa_1^2 \bkappa_3^2\
\bk^{\prime 2} \bkappa_1^{\prime 2} \bkappa_3^{\prime 2}\rightarrow
0$. It has been argued \cite{Durham} that this divergence could be
regularised via the Sudakov form factor, provided one chooses the
smallest gluon momentum as a lower bound for the $\bq^2$ integral in
(\ref{sudasplitting}). This is in general not sufficient, as the
same diagrams lead to a contribution to the inclusive jet production
cross section, which will not contain any Sudakov form factor, but
must be finite nevertheless. Furthermore, is well-known \cite{ChWu}
that the Sudakov form factor is not required to regularise the
divergence: indeed, gluons with a long wavelength $\lambda$ average
the colour of the proton, and hence a suppression proportional to
$r_c/\lambda$ must always be present when coupling to a colour
singlet of size $r_c$. This is usually taken into account via the
introduction of an impact factor $ \Phi(\bk_a,\bk_b)$ such that
\begin{equation}
\Phi(0,\bk_b)=\Phi(\bk_a,0)=0\,.\label{zerocondition}
\end{equation}
Here $\bk_a$ and $\bk_b$ are the transverse momenta of the two 
$t$-channel gluons that couple to the proton.
The differential cross section for two-gluon production in proton-proton 
scattering
can then be represented as
\be
d\sigma_{pp} = d\sigma_{qq} \otimes 
\prod_{i=1,3}\left[\sqrt{T(\bk_2^2,\bkappa_i^2)}\Phi(\bk,-\bkappa_i)\right]
\left[\sqrt{T(\bk_2^2,\bkappa_i^{\prime 2})} 
\Phi(\bk',-\bkappa_i^\prime)\right]\,.
\label{hadronic-xsection}
\ee
The symbol $\otimes$ here indicates that the factors are to be introduced
inside the loop momentum integrals $\int d^2\bk\, d^2 \bk'$.

The presence of the impact factors removes all the infrared singularities 
present
in the partonic-level expressions. Let us also note that $g^{12}$ in 
(\ref{partonic-xsection})
should be understood as $g_{hard}^4\, g_{soft}^8$, where $g_{hard}$ is a 
hard coupling
coming from gluon-gluon collision, while $g_{soft}$ is the coupling at 
the gluon-proton vertex.

Note that when linking hadronic and partonic cross sections as in 
(\ref{hadronic-xsection}),
we do not need to worry about the flux factors and the phase space.
In principle, one could write the cross section for proton-antiproton 
scattering
from the start, which would be expressed in terms of the hadronic fluxes 
and phase space,
and then work out the two-gluon production amplitude in the 
proton-antiproton collision.
The net result
would consist in the replacement of $s$ by $s_{\bar pp}$ both in the 
flux/phase space
and in the amplitude, and since the lowest order cross section is 
independent of the energy,
these changes would cancel each other.
This is consistent with the fact that the partonic lowest-order cross 
section (\ref{partonic-xsection}) depends
on the momentum fractions $\beta_1$ and $\beta_2$ only via
there ratio $x = \beta_1/\beta_2$, so that $\beta_i$
can be taken either with respect to the quark or to the (anti)proton.

This situation changes beyond the lowest order. In particular, we
shall introduce below the unintegrated gluon density of the
(anti)proton that depends on the fraction of the hadronic momentum
carried by the gluon. Therefore, from now on, we shall understand
the fractions $\beta_i$ and $\alpha_i$ with respect to the hadrons,
i.e. as if the decomposition (\ref{ggmomenta}) involved the hadronic
momenta $P^\mu$ and $Q^\mu$, rather than the partonic momenta
$p^\mu$ and $q^\mu$.

\subsection{Form factors from a quark model}
The simplest approach to the quark form factors
is to consider that the ultra-relativistic proton is dominated by a
3-quark Fock state \cite{Soper,Nguyen}.
One can then derive the form factors in terms of
the quark light-cone wave function, and show that the form factor
corresponding to the two gluons of momenta $\bk_a,\bk_b$ coupling to
the same quark is identical to the measured Dirac
helicity-conserving form factor $F_1\left((\bk_a+\bk_b)^2\right)$.
The contribution of the couplings to different quarks can then be
parametrised as $F_1\left(\bk_a^2+\bk_b^2+c\bk_a\cdot \bk_b\right)$:
\be \Phi(\bk_a,\bk_b) = 3\left[F_1((\bk_a+\bk_b)^2) -
F_1\left(\bk_a^2+\bk_b^2+c\bk_a\cdot \bk_b\right)\right],\label{Phi}
\ee This makes the jet production finite. An  $s$ dependence can be easily
introduced, provided one assumes
that a simple-pole pomeron, of intercept $\alpha_0$ and slope
$\alpha'$, is a good approximation. One then has to multiply the
form factor $\Phi$ by the Regge factor~\cite{BL}
\begin{equation}
R_{jets}=1/\xi_i^{\alpha_0+\alpha' t}
\label{shrinkage}
\end{equation}
with $t$ the total momentum carried by the two gluons, and $\xi_i$ the 
longitudinal momentum
losses of the proton or of the antiproton, in our case $\beta_1+\beta_2$ 
or $\alpha_1+\alpha_2$.

The coefficient $c$ can then be fixed, as well as the soft coupling 
$g_{soft}$, so that one reproduces
soft data such as the total cross section and the total elastic cross 
section, for which the Regge factor
has to be changed to
\begin{equation}
R_{soft}=\alpha' s^\frac{\alpha_0+\alpha' t}{2}.
\end{equation}
For the CDF cuts, one has $\xi_i\sim 0.01$, so that
the Regge factor
of the jet cross section
is comparable to that of elastic scattering at 50 GeV. The $t$-slope of 
the exchange is thus of the
order of 4 to 5 GeV$^{-2}$.

The two main disadvantages of this method come from the fact that, even 
if simple-pole exchanges dominate the soft amplitudes up to the Tevatron 
energy, they will receive non-negligible corrections at LHC energies. 
Furthermore, the use
of perturbative 2-gluon exchange to reproduce elastic cross sections does 
not work, in the sense that
it produces a very large curvature for $d\sigma/dt$. So the normalisation 
of $c$ and $g_{soft}$ in  this approach
is at best tentative.

It may thus make sense to use DIS data to normalise the impact factors. 
This is precisely the
idea behind the use of unintegrated gluon densities in this calculation.

\subsection{Unintegrated gluon density}\label{subsec-embedding-ugd}

The distribution of gluons inside an ultra-relativistic proton can be modelled
in a diagonal process by replacing the Born-level impact factor of the 
quark model
with an unintegrated gluon density \cite{IN2000}:
\be
C_F {g^2\over 4\pi^2} \Phi(\bk,-\bk) \equiv {\cal F}_{\mathrm{Born}}(x_g,\bk)
\to {\cal F}(x_g,\bk)\,.
\label{uninteglu}
\ee
In numerical calculations we used parametrisations of the unintegrated
gluon density developed in \cite{IN2000,IvanovPHD}.
These parametrisations were obtained by fitting, in a $k_t$-factorisation 
approach, the proton structure function $F_{2p}$ to the HERA data
in the region of photon virtuality $0< Q^2 < 35$ GeV$^2$ and for Bjorken-$x$
smaller than $10^{-2}$.
Although the $k_t$-factorisation approach is designed mostly to work in
the region of small $x$ and moderate $Q^2$,
it was verified that the inclusion of a valence quark contribution
extended these fits of $F_{2p}$ data to a much broader kinematic
region,
for $Q^2$ up to 800 GeV$^2$ and for $x$ up to $\sim 0.5$.
These fits, without any readjustment, were also found to
give a good description of the charm contribution to $F_{2p}$, $F^{c}_{2p}$,
of the longitudinal structure function, $F_{L}$, as well as of diffractive
vector meson
production \cite{IvanovPHD,INS2005}. This serves as an important
cross-check of the universality of the unintegrated distributions.

These fits of ${\cal F}(x_g,\bk)$ were constructed as sums of two terms,
a hard part and a soft part, with a smooth interpolation between them.
The hard component describes the effects of hard perturbative gluon
exchange, and therefore is based on direct differentiation
of conventional gluon densities
(for which LO fits by GRV\cite{GRV}, MRS\cite{MRS}, and CTEQ\cite{CTEQ} 
were used)
and smoothing out the saw-like behaviour of the result.
The soft part describes soft colour-singlet exchange
in the non-perturbative regime,
and is constructed in a phenomenological way inspired by the dipole
form factor.
We stress that in the soft regime the term ``unintegrated gluon density''
must be understood simply as the Fourier transform of the colour dipole
cross section.

One also should not forget that at small $x$ there is no strong hierarchy among
successive gluons in a $t$-channel gluon ladder, which implies
that the boundary between soft and hard
interactions becomes very smooth.
For example, because of this soft-to-hard diffusion,
the robust feature of all the parametrisations obtained in
\cite{IN2000,IvanovPHD}
was that at $x\sim 10^{-2}$, the structure function $F_{2p}$
received dominant contributions from the soft part for photon virtualities
up to $Q^2\sim 10$ GeV$^2$.
Since the process we consider in this paper also takes place at small $x_g$,
this brings further concerns on the validity of the often-cited
statement that the central exclusive diffractive production
is in the perturbative regime and is well under control.

One of the manifestations of the soft dominance in this kinematic
region is the observation that  exponent $\lambda$ which controls
the energy growth
of the integrated gluon density
(often called  the effective Pomeron intercept) calculated within
these fits is significantly smaller than the one calculated directly from
the conventional gluon densities.
At typical gluon transverse momenta of 1--2~GeV$^2$, we find
$\lambda \sim 0.1-0.2$, instead of $0.3-0.4$, as obtained from the
conventional gluon densities.
This is expected since the standard DGLAP approximation
does not take into account new portions of the phase space
that open up at small $x$, and instead attributes
an artificially fast growth rate to the gluon density itself.

The structure function $F_{2p}$ allows one to obtain the fits of the
forward unintegrated gluon density, while the process we study here contains
skewed unintegrated gluon densities at non-zero momentum transfer
${\cal F}(x_1,x_2,\bk_a,\bk_b)$.
We construct the latter in the following way.
The effect of skewness is effectively taken into account
by assuming that, for  $x_1 \ll x_2$, ${\cal F}(x_1,x_2,\bk_a,\bk_b)$ 
behaves as the forward
density taken at $x_g=0.41 x_2$.
In our case, $x_2=\beta_1+\beta_2$ for the upper and $x_2=\alpha_1+\alpha_2$
for the lower proton.
The coefficient 0.41 effectively takes into
account the skewness factor, introduced in \cite{skew} and used in 
\cite{Durham}.
Numerically, this correction amounts to a factor $0.41^{-4\lambda} 
\approx 1.2^4 \approx 2$.
Calculation of this factor using the conventional gluon densities
would give a much larger value, $1.4^4 \approx 4$.

The non-zero transverse momentum transfer was introduced via
a universal exponential factor constructed in such a way that it respects
conditions (\ref{zerocondition}) and takes shrinkage into account,
similarly to \cite{IvanovPHD,INS2005}. For example, the full expression
for the upper proton used in our calculations reads:
\bea
{\cal F}(x_1,x_2,\bk,\bkappa_1) &=& {\cal 
F}\left(0.41(\beta_1+\beta_2),{\bk^2+\bkappa_1^2 \over 2}\right)\nonumber\\
&&\times {2\bk^2\bkappa_1^2 \over \bk^4+\bkappa_1^4}
\exp\left\{-{1 \over 2}\left[B_0 + 2\alpha'\log\left({x_0 \over 
\beta_1+\beta_2}\right)\bk_1^2\right]\right\}\,,
\label{offforward}
\eea
with $B_0 = 4$ GeV$^{-2}$, $\alpha'=0.25$ GeV$^{-2}$ and $x_0 = 3.4\cdot 
10^{-4}$.
With these values of the diffractive cone parameters, the slope of
the $\bk_1^2$ distribution
$d\sigma/d\bk_1^2 \propto \exp(-B_p\bk_1^2)$ is approximately equal to
$B_p \approx 4$ GeV$^{-2}$.
This is significantly smaller than the slope in the elastic $p\bar p$ collision at the Tevatron energy,
$$
{d\sigma_{pp}(el.) \over d|t|}\ \propto\ \exp(-2 B_{el} |t|)\,,\quad B_{el} \approx 8.5\, \mathrm{GeV}^{-2}\,.
$$
In our opinion, the choice $B_p \approx 4$ GeV$^{-2}$ is more natural 
than $B_p \approx B_{el}$
since the elastic scattering at the Tevatron energy probes gluon 
distribution at $x\sim 10^{-7}$,
while in our process the energetic gluons carry $\sim 10^{-2}$ of the 
proton momenta.
However we stress that this choice is model dependent, and it introduces 
an extra
uncertainty into theoretical calculation of the central production cross 
sections.

Finally, note that we incorporate the unintegrated gluon density and the 
Sudakov form factor
as independent factors. It might be argued \cite{CCFM2,kwie}
that a more correct procedure would be to {\em define}
the unintegrated gluon density via a derivative of conventional gluon 
density times the square
root of the Sudakov form factor. Either way one obtains only
a convenient parametrisation of the true unintegrated gluon distribution 
function.
The only essential requirement is that the same prescription be used for
all the processes. We believe that the way our fits were constructed in 
\cite{IN2000}
and are implemented here, this requirement is satisfied.

\section{Gap survival}\label{sec-gapsurvival}

There is one final aspect that has to be tackled to finish the
calculation, and it has to do with gap survival probability. This
concept was introduced by Bjorken a long time ago \cite{BJ}, and has been 
recently
investigated in detail in \cite{Frankfurt}. We remind the reader of the main
argument. As shown in Fig.~\ref{fig:steps}, the process that we have 
calculated at the proton
level may still have to be corrected for initial and final-state interactions.
The argument is that, due to the fact that the hard interaction occurs at short
distance, and does not change the quantum numbers of the protons, it does
not influence the rescatterings.
Of course, these can change the transverse momenta
of the protons, and one would have to convolute the hard scattering with 
the multiple exchanges.
However, these correlations disappear once one works in impact parameter space,
so that the total interaction probability can be thought of as the 
product of the hard scattering probability
multiplied by the probability for the two protons to go through each 
other, {\it i.e.}
the $S$ matrix element squared $|S(\bb)|^2=|\langle pp| S | 
pp\rangle|^2$. It is easy to work out
the square of the absolute value of this correction from the expressions of the
total and of the elastic cross sections.

Starting with
\begin{equation}
\frac{d\sigma}{dt}=\frac{1}{16\pi s^2} |a(s,t)|^2
\end{equation}
one can use the usual definition of $S$
\begin{equation}
S(\bb)=1+ia(s,\bb)\label{gapsu}
\end{equation}
to get  the partial wave
\begin{equation}
a(s,\bb)=\int\frac{d^2\mathbf\Delta}{(2\pi)^2}\frac{a(s,t)}{2s}
\end{equation}
which leads to the expressions
\begin{eqnarray}
\sigma_{tot}&=&2\int d^2\bb\ \mathrm{Im}\ a(s,\bb),\\
\sigma_{el}&=&\int d^2 \bb\ \left|a(s,\bb)\right|^2.
\end{eqnarray}
The square of the $S$-matrix density is then the square of the deviation
of $a(s,\bb)$ from $i$
\cite{Troshin,Frankfurt}:
\begin{equation}
|S(\bb)|^2=|i-a|^2
\end{equation}
Hence any fit of the differential elastic cross section can be used
to estimate the gap survival probability. Generically, the gap survival
will tend to 1 at large $\bb$ and be suppressed at small $\bb$.
All present models agree
on the fact that, at the Tevatron, the elastic amplitude approaches
the black-disk limit $a(s,\bb)=i$ for small $\bb$.

The simplest fit of the elastic cross section comes from CDF, who fit 
their data to \cite{CDFelastic}
\begin{equation}
\frac{d\sigma}{dt}=N\, \exp\left(2B_{el} t \right)
\end{equation}
with $N=334.6\pm 18.8$ mb GeV$^{-2}$ and $2B_{el}=16.98\pm0.24$ GeV$^{-2}$.
From this, one can get an estimate of the gap survival probability,
by assuming that the amplitude is purely imaginary. One then gets
\begin{equation}
a(s,\bb)\approx\frac{i}{2B_p}\sqrt{\frac{N}{\pi}}\exp
\left(-\frac{b^2}{4B_{el}}\right)=
i(0.974\pm 0.042)\exp\left(-\frac{b^2}{4B_{el}}\right)\label{gapb}
\end{equation}
Clearly, this is very close to the black-disk limit at $b=0$. So one may 
expect a substantial suppression of the
cross section due to screening corrections. If the cross section is at 
really short distance, then the average
gap survival would be around $|S(0)|^2$, i.e. less than 0.5 \%.

Fortunately, this estimate is overly pessimistic as the cross section is 
not concentrated at very short distance.
Moreover, our estimate (\ref{gapb}) gives only a lower bound, the 
amplitude has a real part, and
also as the measurements of the E811 collaboration \cite{E811}
suggest one may be further from the black disk limit. Furthermore, 
although the differential cross section is close
to an exponential near $t=0$, it has another structure at higher $t$. 
Putting all these ingredients together
is beyond the scope of this paper. Let us simply mention that several 
estimates are present in the literature \cite{pancheri},
and that most of them range from 5\% to 15\% at the Tevatron, and about a 
factor 2 lower at the LHC.

We want however to point out a few problems with the standard 
calculations of gap survival:
\begin{itemize}
\item the conjugate variable of $\bb$ is $\bk_1+\bk_3$, the relative 
momentum of $p$ and $\bar p$. This is
a combination of soft momenta, and folding in
the gap survival (which is small at small~$\bb$) will further shift these 
momenta to
the long-distance region. It
is then very unlikely that the screening corrections will be given by the 
simple gap
survival formalism. In fact the screening corrections will probably
have a smaller effect, as they do not suppress truly soft
cross sections very much at the Tevatron.
\item most estimates \cite{pancheri} assume that the $\bk_1$ and $\bk_3$ 
dependences factorise. This makes the calculation
of the gap survival quite simple. However, because of the $\bk$ and 
$\bk'$ integrations, this is not true in our case.
Given the large uncertainties in the gap survival, this may not be a 
crucial issue.
\end{itemize}

\section{Rough Estimate of the Cross
Section}\label{sec-roughestimate}

\subsection{Bare cross section}

Before presenting detailed numerical results, we find it useful to make
some simple order-of-magnitude estimates.

Let us start by estimating the bare cross section at the hadronic level,
i.e. by keeping the proton form factors but omitting the Sudakov form 
factor altogether.
Consider again the cross sections (\ref{partonic-xsection}), 
(\ref{hadronic-xsection}).
Since the strongest dependence on the $\bk_1$ and $\bk_3$ comes
from the proton form factors, it is reasonable to take 
$\bkappa_1\approx\bkappa_3\approx\bk$ and
$\bkappa_1'\approx\bkappa_3'\approx\bk'$
in the numerator. Assuming the correlation between $\bk$ and $\bk^\prime$ 
to be weak,
one can neglect the total helicity 2 amplitudes, so that
the cross section simplifies to
\be
d\sigma = {1 \over 9 \pi} {d\bk_2^2\over(\bk_2^2)^2}\, {d\beta 
\over\beta} {dx \over x}
\alpha_{h}^2\  \alpha_{soft}^4 \  \langle \Phi^4\rangle .\label{dsigma2}
\ee
with $\alpha_{hard}=g_{hard}^2/4\pi$ and correspondingly for $\alpha_{soft}$.
 We introduced here a dimensionless quantity
\be
\langle \Phi^4\rangle \equiv {1 \over\pi^4}\int d^2\bk_1\, d^2\bk_3
\left[\int {d^2\bk\ \Phi(\bk,-\bk-\bk_1) \Phi(-\bk,\bk+\bk_3)\over 
(\bk+\bk_1)^2(\bk+\bk_3)^2}\right]^2\,.\label{Phi4}
\ee
Note that the $x$-integration in (\ref{dsigma2}) spans over the $x\ge 1$ region, which allows us not to include
the $1/2$ factor due to the Bose statistics of gluons.
The longitudinal phase space integration gives
\be
L \equiv \log(\beta_{\mathrm{max}}/\beta_{\mathrm{min}})\  \log 
x_{\mathrm{max}}\,.\label{Delta}
\ee
The cross section integrated over $\bk_2^2>\bk^2_{2, min}$ is
\be
\sigma \approx {1 \over\bk^2_{2, min}}\ {1 \over 9\pi} \ L \  
\alpha_{hard}^2\alpha_{soft}^4
\  \langle \Phi^4\rangle,\label{estimate1}
\ee
which gives for $|\bk_{2, min}|=10$ GeV
\be
\sigma \approx 0.14\, \mu\mathrm{b}\ L \  \alpha_{hard}^2\alpha_{soft}^4
\  \langle \Phi^4\rangle .\label{estimate2}
\ee

Inserting the cuts used by Tevatron, we get $L \approx {\cal O}(5)$,
while $\alpha_{hard}^2 \sim 0.04$.
Then, one has to estimate $\langle \Phi^4\rangle$, which is an 
intrinsically soft quantity.
Let us recall the expression of the elastic $pp$ scattering in the same 
approximation:
\be
\sigma_{el} =  {4\alpha_{soft}^4 \over 81\pi^2}
\int d^2\bk_1 \left[\int {d^2\bk\ \Phi^2(\bk,-\bk-\bk_1) \over \bk^2 
(\bk+\bk_1)^2}\right]^2\,,
\ee
Assuming that the strongest $\bk_1$ dependence comes from the exponential 
diffractive factor
in $\Phi$ and neglecting the energy dependence of $\Phi$, one can roughly 
estimate
\be
\langle \Phi^4\rangle \sim {1 \over \pi^2 B_p^2}
\left[\int {d^2\bk\over \bk^4}\Phi^2(\bk,-\bk) \right]^2\,,
\quad \sigma_{el} =  {2\alpha_{soft}^4 \over 81\pi B_p}
\left[\int {d^2\bk\over \bk^4}\Phi^2(\bk,-\bk) \right]^2\,,
\label{estimatephi}
\ee
so that
\be
\alpha_{soft}^4\, \langle \Phi^4\rangle \, \sim {81 \over 2\pi 
B_p}\,\sigma_{el}
\sim {\cal O}(100).
\ee
The estimate (\ref{estimate1}) now reads:
\be
\sigma \approx {9\over 2\pi^2 \bk^2_{2, min}}
\,{\sigma_{el} \over B_p} \, L\ \alpha_{hard}^2 \approx {1 \over 
\bk^2_{2, min}}\,.
\label{estimate3}
\ee
For $|\bk_{2}|_{min}=10$ GeV, it gives very roughly $\sigma \sim  {\cal 
O}(4\ \mu b)$.

This estimate, which normalises the jet cross section to the elastic 
cross section, includes implicitly
a gap survival probability,
which we assume here to be of the same order of magnitude for both processes.

\subsection{Sudakov suppression}\label{subsec-sudakov-estimate}

CDF has measured the dijet central exclusive cross section to be about 1 
nb at \hbox{$E_{T}^{min}=10$ GeV,} which is
three orders of magnitude below the above estimate. It indicates that 
the Sudakov suppression
indeed plays a crucial role in this process.

The Sudakov form factor enters the loop integral, and it reshapes
the $\bk^2$ regions that contribute most to the amplitude. Before
the introduction of the Sudakov form factor, the loop was dominated
by the soft momenta due to $1/\bk^4$ factor. Now, the weight of the
soft momenta is suppressed in Eq.~(\ref{estimatephi}):
\be {\cal J}=\int {d\bk^2 \over
\bk^4}\,\Phi^2(\bk,-\bk)e^{-S(\bk_2^2,\bk^2)}\,.\label{estint}
\ee
As a result, the dominant $\bk^2$-region shifts towards harder
scales. Using the saddle point approximation, it has been estimated
\cite{Durham} that the dominant region is at $|\bk| \approx$ 1--2
GeV. It is often claimed that this makes the loop sufficiently hard
to justify the applicability of pQCD and the usage of perturbative
fits to the gluon density.

Here, we discuss this issue in some detail.
With two simple estimates, we will show below that the overall 
suppression and the $|\bk|$ shift
depend strongly on the details of the Sudakov form factor.

Let us switch to a more convenient variable $x = \log(\bk^2/\Lambda^2)$,
where $\Lambda = \Lambda_{QCD}$ and rewrite (\ref{estint}) as
\be
{1 \over \Lambda^2} \int dx \,e^{-{\cal I}(x)}\,,\quad
{\cal I}(x) = x - 2 \log\Phi(x) + S(x_2,x)\,,\label{Ixdef}
\ee
where $x_2 \equiv \log(\bk_2^2/\Lambda^2)$.

In the first approximation, the proton form factor $\Phi$ plays the role 
of the infrared cut-off
of the above integral: $x\gsim 0$. Therefore, without the Sudakov form 
factor altogether,
the integral in (\ref{Ixdef}) is of order one.

Now, let us take the Sudakov form factor in the double-log approximation 
(\ref{sudakDLL1})
with the fixed $\alpha_S(\bk_2^2)$.
Then,
\be
{\cal I}(x) = x + {3 \over \beta_0 x_2}(x_2 - x)^2\,,
\ee
and the position of the saddle point $x_0$ is at $x_0 = x_2 (1 - 
\beta_0/6) < 0$, i.e.
the integral is still saturated in the soft region. The estimate of the suppression
factor is then
$$
\int_0^\infty dx \, \exp\left[-{\cal I}(0)-{\cal I}\,'(0)\,x\right] =
{\exp\left[-3 x_2/\beta_0\right] \over 1-{6 \over\beta_0}}\,,
$$
which is about 0.2 for $k_2=$ 10 GeV.
If, instead, we take a running $\alpha_S(\bk^2)$, then in the same 
double-log approximation
\be
{\cal I}(x) = x + {6 \over \beta_0}\left(x_2\log{x_2 \over x} -  x_2 + 
x\right)\,, \label{Ixrunning}
\ee
which gives now
\be
x_0 = {x_2 \over 1 + {\beta_0 \over 6}}\,,
\ee
which now makes typical $\bk^2 \sim$ 1 GeV. The integral then becomes
\be
\sqrt{{\pi \beta_0 x_2\over 3 (1+{\beta_0\over 6})^2}}
\exp\left[-{6 x_2\over\beta_0}\log\left( 1+{\beta_0\over 6}\right)\right],
\ee
which is about $8\cdot 10^{-3}$ for $k_2=$ 10 GeV.

Such a severe qualitative change of the result clearly indicates its
sensitivity to different prescriptions used in the Sudakov form
factor. In the view of additional single-log corrections coming from
unspecified scales in the logarithms as well as from the extra
diagrams discussed in Section~\ref{subsec-screening}, which are not
yet under control, one must conclude that the claims of
perturbativity of the $\bk$-loop are unjustified. Depending on the
assumptions used, the Sudakov suppression can differ by one order of
magnitude, and mean $\bk^2$ can vary from ${\cal O}(0.1)$ to ${\cal
O}(1)$ GeV$^2$. This uncertainty affects not only numerical results,
but also some qualitative arguments.

Nevertheless, we see that it is possible to get a suppression factor of 
the order of 100 from the virtual corrections.
This reduces our estimate to about 10 nb.
An extra suppression of a factor 3 comes from
the shift in $E_T$ when one goes from partons to jets. We shall discuss 
this in the next section. So we
see that all the ingredients of this calculation can lead to a cross 
section in rough agreement with the data.
In fact, we shall show that it is possible to get an exact agreement with 
the data, but that the number of
adjustable theoretical corrections allows for many theoretical 
possibilities. Hence the CDF data turn out
to be very important for tuning the theory.

\section{Numerical results}\label{sec-numerical}
\subsection{Cuts}

The two gluons produced through pomeron exchange hadronise into
jets. As is unavoidable, this hadronisation can result in any number
of jets, not just two. Theoretically, the difference between the two
gluon cross section and the dijet cross section ({\it i.e.} the 3-jet
veto) is a correction of the order of $\alpha_S(E_T)$
\cite{durjets}. Given the large theoretical uncertainties in other
parts of the calculation, we do not take this correction into
account.

On the other hand, the transition from the partonic to the hadronic
level involves corrections which are larger, and which are due to
radiation outside the jet-finding cone --- we shall generically
refer to them as splash-out. The structure of these corrections is
known \cite{stirling,salam} and involves a constant shift in $E_T$
of the order of 1 GeV, due to hadronisation, as well as a correction
proportional to $\alpha_S(E_T) E_T$ due to radiation. The
corresponding shift in $E_T$ has been estimated, for the cone
algorithm used by CDF \cite{splashout2} to be
\begin{equation}
E_T^{jet}=(0.75-0.80) k_T^{parton}
\label{splasheq}
\end{equation}
 We shall also consider a previous parametrisation \cite{splashout1}, where
\begin{equation}
E_T^{jet}=k_T^{parton}\left[1-{1\over 2}\ 
\alpha_S\left({(k_T^{parton})}^2\right)\right] -1\ {\rm GeV}.
\end{equation}
One can then compare our results to the data, for which the cuts are 
summarised in Table~\ref{cuts}.
\begin{figure}
\begin{center}
\epsfig{file=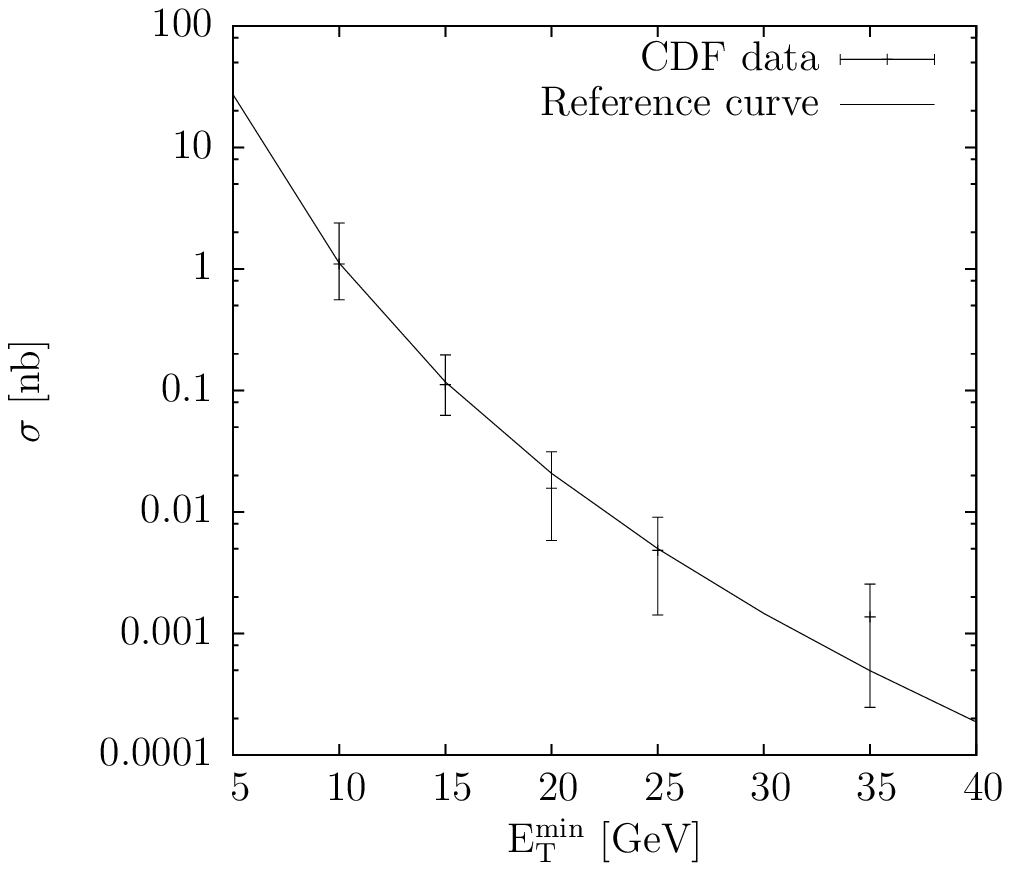,width=8cm}
\epsfig{file=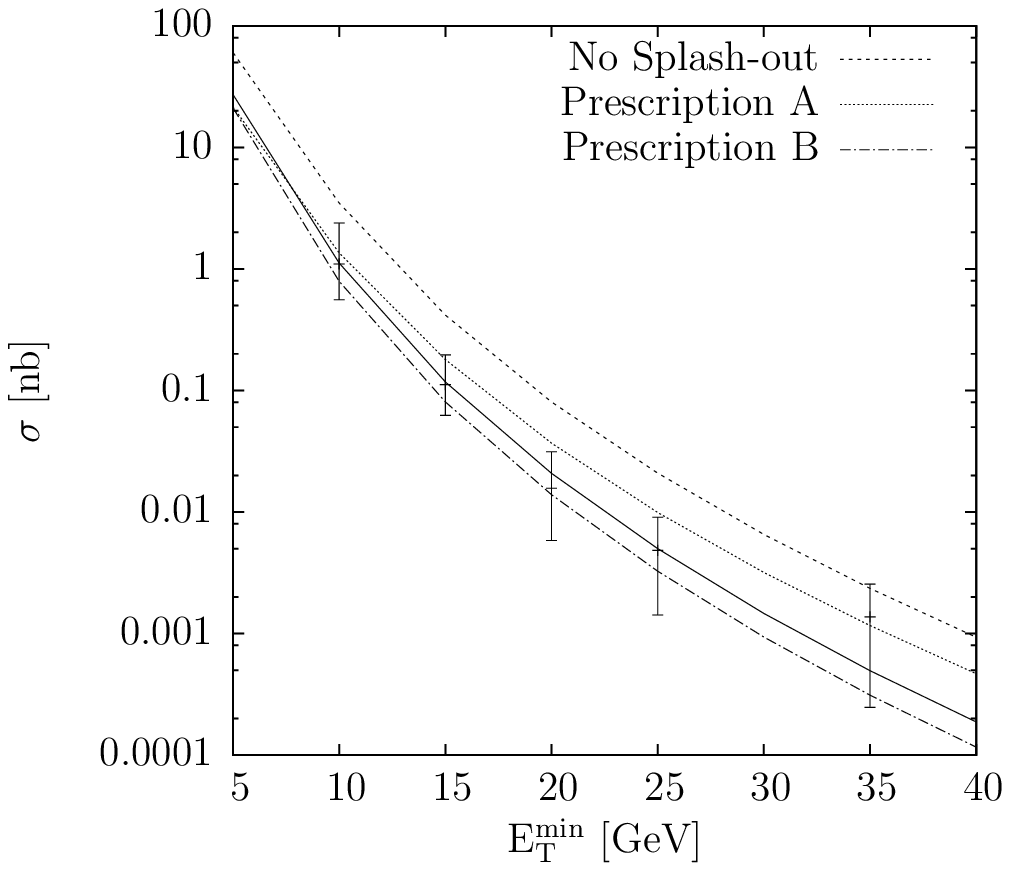,width=8cm}\\
~\hglue 0.5cm(a)\hglue 7.5cm (b)\\
\caption{(a) Our reference curve, corresponding to the parameters of
Table~\ref{default}, chosen so that it goes through the CDF run II
data \cite{CDF}. (b) The suppression of the cross section for
various splash-outs discussed in the text. The plain curve is as in
(a).\label{referencefig}}
\end{center}

\end{figure}
\begin{table}
\begin{center}
\begin{tabular}{|c|c|}
\hline  $\alpha_1+\alpha_2$& [0.03, 0.08] \\
\hline  
$y_{jet}^{(i)}=\frac{1}{2}\log\left(\frac{\beta_i}{\alpha_i}\right)$, 
$i=1,$ $2$& [$-$2.5,2.5] \\
\hline  $|y_p-y_{jet}^{(i)}|$, $i=1,$ $2$&  $>3.6$\\
\hline  $|y_{\bar p}-y_{jet}^{(i)}|$, $i=1,$ $2$&$>3.6$\\
\hline
\end{tabular}
\end{center}
\caption{Experimental cuts\label{cuts} for CDF run II data \cite{CDF}.}
\end{table}
We show in Fig.~\ref{referencefig}.b the effect of these corrections. We 
see that for the $E_T$ range of the CDF data, the effect of
splash-out is non-negligible\footnote{We thank V. Khoze for pointing this 
out to us.} and amounts to a correction of the
order of a factor of 3 for $E_T^{min}=10$ GeV, and the various 
possibilities for the splash-out bring in an
uncertainty of the order of  1.7 at $E_T^{min}=10$ GeV to 4 at 
$E_T^{min}=35$ GeV. Note that the largest effect comes
from the shift in $E_T^{min}$. We have also considered a smearing in the 
rapidity of the jets, of the order of 1 unit, but this has
an effect on the cross section of less than 0.1\%. So it seems that in 
the longitudinal direction, the cuts of Table \ref{cuts}
can be directly applied at the parton level.

In the following, we shall fix the splash-out correction to
$E_T^{jets}=0.80 E_T^{partons}$. As a reference we shall also choose
the parametrisation given in Table \ref{default}. This is not
necessarily our best guess, but rather one of the choices which
reproduces the CDF data. We shall include other possibilities when
we extrapolate our results to the LHC.

\begin{table}
\begin{center}
\begin{tabular}{|c|c|c|}
\hline  parameter& value&equation \\\hline
\hline $\Lambda_{QCD}^{(5)}$&0.20 GeV & ~\\
\hline  scale of $\alpha_S$ in partonic cross section&  $s_{gg}$& \\
\hline  scale of $\alpha_S$ in Sudakov form factor& $\bq^2$& 
(\ref{sudasplitting})\\
\hline angular ordering&yes&$\Delta = |\bq|/\mu$ in 
(\ref{sudasplitting})\\
\hline  terms in Sudakov exponentiation& constant+log+log$^2$ & 
(\ref{sudasplitting})\\
\hline  lower scale of Sudakov integral& $\ell^2=(\bk+\bk_i)^2$& 
(\ref{sudasplitting})\\
\hline  upper scale of Sudakov integral&  
$\mu^2=\bk_2^2/2$&(\ref{sudasplitting})\\
\hline  unintegrated structure function& ref. \cite{INS2005}& 
(\ref{uninteglu})\\
\hline  gap survival probability&  $\langle S^2\rangle =15 
\%$&(\ref{gapsu})\\
\hline  splash-out&  $E_T^{jet}=0.8 E_T^{partons}$&(\ref{splasheq})\\
\hline
\end{tabular}
\caption{Default parameters of the reference curve of
Fig.~\ref{referencefig}.\label{default}}
\end{center}
\end{table}

\subsection{Properties of the amplitude}\label{subsec-properties}
The accuracy of several properties and approximations presented in
the literature can be directly tested from their effect on the cross
section.
\begin{figure}
\begin{center}
\epsfig{file=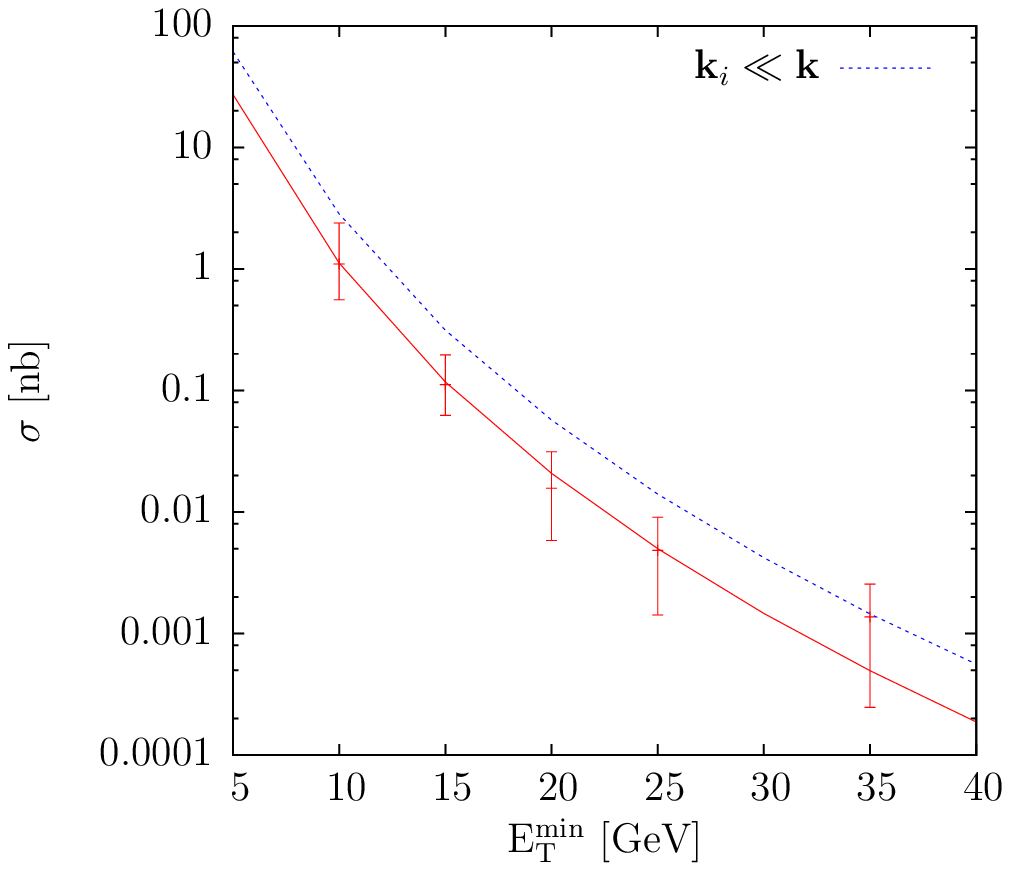,width=8cm}
\epsfig{file=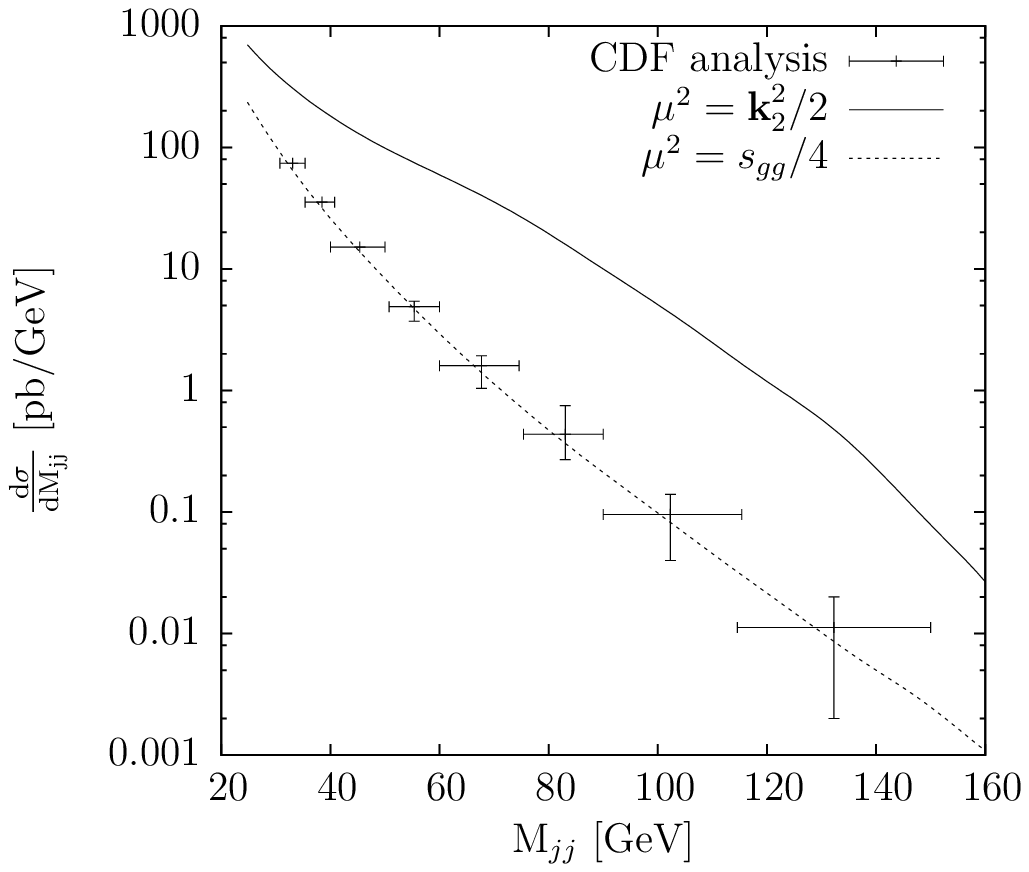,width=8cm}\\
~\hglue 0.5cm(a)\hglue 7.5cm (b)\\
\caption{(a) The effect of
neglecting $\bk_i$ w.r.t. $\bk$ or $\bk'$ with the parameters of Table 
\ref{default}.
(b)  Mass distribution of the jet system, compared with CDF run II data 
\cite{CDF}, when changing the upper scale of the Sudakov form factor, as 
indicated.  In both figures the plain curve is as in 
Fig.~\ref{referencefig}.\label{nonpert}}
\end{center}
\end{figure}

The first property is the claim that the cross section is perturbative.
We define $\sigma_{pert}$ as the value of the cross section in which all 
gluon momenta are larger
than 1 GeV.
Although this is not a physical observable, it is a useful quantity to test the
assumptions used in theoretical calculations.
We find that at CDF, the ratio $\sigma_{pert}/\sigma$ is 0.35 for 
$E_T^{min}=10$~GeV, and falls to 0.25 for $E_T^{min}=35$~GeV.
Clearly, as already
mentioned, the effect of the Sudakov form factor is not sufficient to 
shift the dominant values of the momenta into the
perturbative region, and we see that momenta below 1 GeV contribute 
significantly to the cross section.
Note also that the larger $E_T^{min}$, the {\em softer} the gluon
loop. This is expected because at fixed proton energy, the
production of higher-$E_T$ jets requires larger $\alpha_i$ and
$\beta_i$, and it is a robust property of the unintegrated gluon
distributions that they become softer at larger $x$, see
e.g.\cite{IN2000}.

It might be, under
those conditions, that modifications to the gluon propagator have to be 
taken into account \cite{Bzdak}. It remains
largely unclear however how to take these into account in a consistent 
and gauge-invariant way, as modifications to the
propagator should also involve modifications to the vertices. Also, we 
have not taken into account
the effect of the gap survival factor, which will reduce the dominant 
values of $\bk_i$. Again, the formalism to follow seems
unclear as the protons communicate with the jet via rather soft momenta.

The second point we can test is the neglect of the momenta $\bk_i$ in 
front of $\bk$, which we used in our
order-of-magnitude estimate (\ref{estimatephi}).
We show in Fig.~\ref{nonpert}.a that this approximation is very rough,
and that it overestimates the cross section at high $E_T^{min}$.

Finally, we confirm that the
 $\lambda=0$ terms of the cross section (\ref{partonic-xsection}) 
dominate the $\lambda=2$ contributions,
 as noticed in \cite{Durham}. The
latter contribute 2\% at $E_T^{min}=10$~GeV to 3\% at $E_T^{min}=35$~GeV.

CDF also presented in \cite{CDF} the dijet mass $M_{jj}$ distribution of 
the jet system.
Note that these are not true data but rather predictions of ExHuMe Monte 
Carlo simulations \cite{Exhume}
normalised to the data: the ExHuMe cross section is normalised to the 
data in a given $E_T$ bin, and
the $M_{jj}$ distribution comes from summing over all these $E_T$ bins. 
The data is presented for $M_{jj}>30$ GeV,
i.e. where the lower $E_T$ cut has little influence.
In Fig.~\ref{nonpert}.b, we show how our calculations compare with these 
distributions. We have produced two
curves, both corresponding to minimum $E_T$ of 5 GeV, and assumed that 
the longitudinal splash-out is the same as the transverse one,
{\it i.e.} we took $M_{jj}=0.80 \sqrt{s_{gg}}$. The first curve 
corresponds to our reference curve.
It clearly overshoots the CDF points, and predicts a considerably harder 
$M_{jj}$ spectrum than the ExHuMe Monte Carlo.
The second curve goes perfectly though the points. The only difference is 
that in the latter case, the upper
scale of the Sudakov form factor has been taken as $\mu^2=s_{gg}/4$, 
close to the choice of ExHuMe
which takes $\mu^2=s_{gg}/2.62$, whereas in the former case that scale 
was $\bk_2^2/2$.
We see that the two choices of the scale lead to vastly different result, 
as they affect the dependence in
$x=\beta_1/\beta_2$ of the cross section.

We argued in Section~\ref{subsec-upperscale} that the upper scale in the 
Sudakov integral
for the dijet production should be related with the relative transverse 
momentum rather than the invariant mass of the dijet.
Thus, $d\sigma/dM_{jj}$ obtained from experimental data without the 
theoretical bias just described
would help test this point.

\subsection{Uncertainties}\label{subsec-uncertainties}
So far, we have shown that it is possible, with an appropriate choice of 
parameters, to reproduce the CDF dijet data
via a calculation containing several perturbative ingredients,
although the dominant momenta are largely in the non-perturbative region. 
We shall now
see why this is the case: there is no firm reason to believe the 
parameters of Table \ref{default}, and different reasonable
choices can easily lead to factors of a few up or down. Basically, all 
the lines of Table~\ref{default} can be changed
to check on the resulting variation. We shall mention here only the most 
significant ones.
We stress here that we are not trying to get the highest possible factor: 
our estimates are
rather conservative, and based on changes that modest theoretical changes 
bring into the
calculation.

\begin{figure}
\begin{center}
\epsfig{file=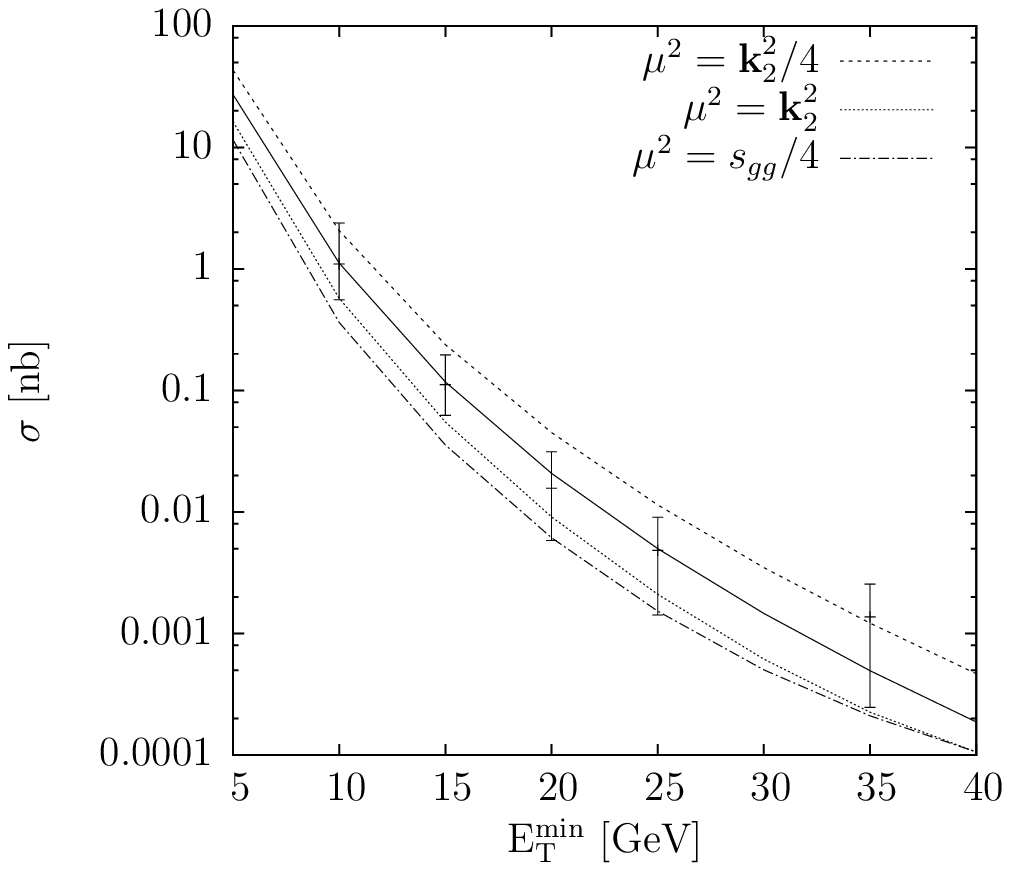,width=8cm}
\epsfig{file=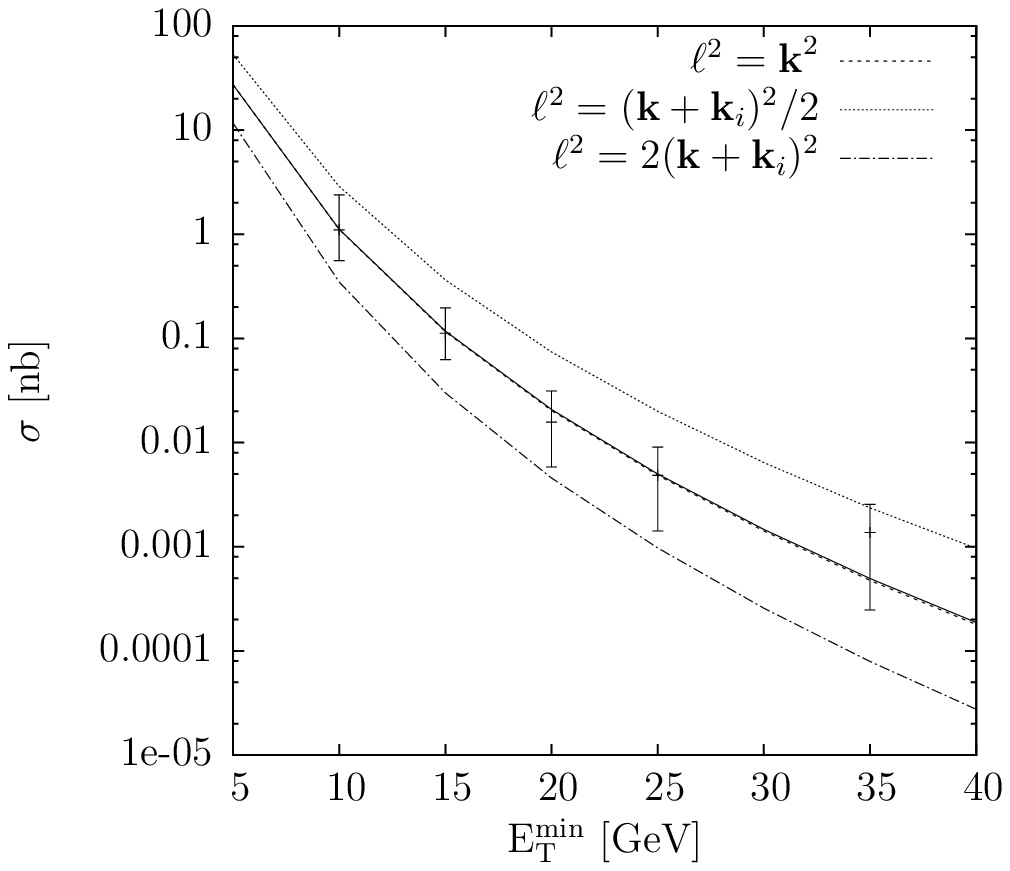,width=8cm}\\
~\hglue 0.5cm(a)\hglue 7.5cm (b)\\
\caption{(a) The effect of changing the upper limit of the Sudakov loops 
by a factor 2, or of choosing $s_{gg}/4$
instead of $\bk_2^2$;
(b) The effect of changing the lower limit of the Sudakov loops by a 
factor 2.\label{bound} In both figures,
the plain curve is as in Fig.~\ref{referencefig}.}
\label{fig-sudabounds}
\end{center}
\end{figure}

The main correction comes from the inclusion of vertex corrections in the 
form of a Sudakov form factor.
The upper and lower limits of the integral (\ref{sudasplitting}) can be 
modified while keeping single-log accuracy.
We show in Fig.~\ref{fig-sudabounds} the effect of such modifications.

As we explained above, these plots clearly show that, in this region of 
$E_T$, the vertex corrections lead
to a rather uncertain situation. On the one hand, they are needed to 
reproduce
the data, but on the other, their numerical impact depends on the details
of their implementation.
We estimate the uncertainty coming from variations of the limits of 
integration as about a factor 3 for the lower limit, and 6 for the upper 
one.

\begin{figure}
\begin{center}
\epsfig{file=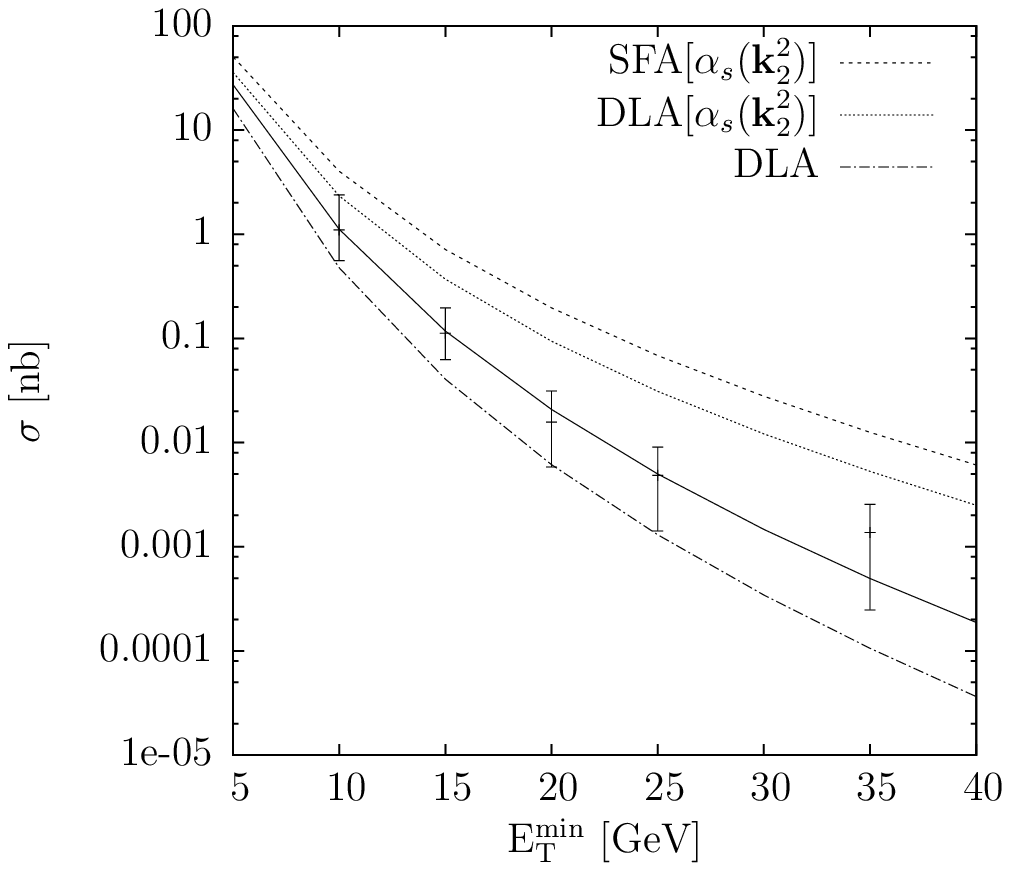,width=8cm}
\epsfig{file=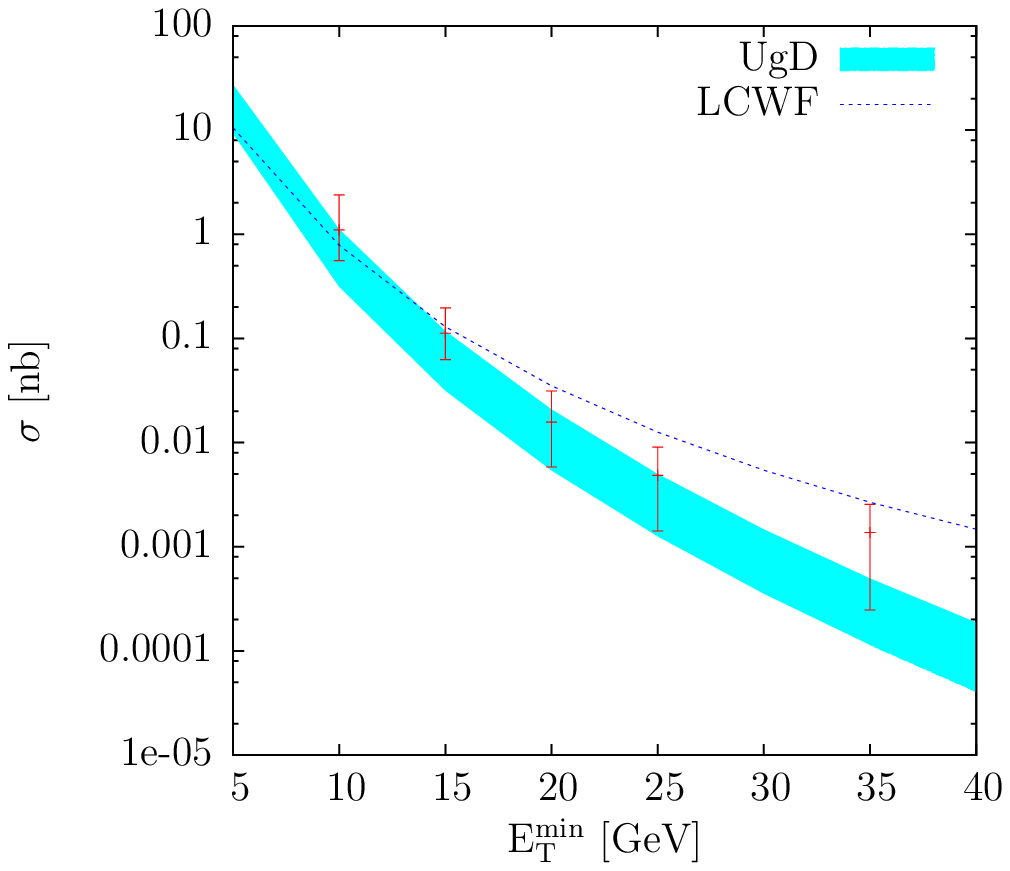,width=8cm}\\
~\hglue 0.5cm(a)\hglue 7.5cm (b)\\
\caption{(a) The differential cross section for different choices of the 
Sudakov form factor: DLA stands for the double log approximation, SFA for 
the splitting function approximation, as defined in 
Section~\ref{subsec-sudakov-resum}. The plain curve is as in 
Fig.~\ref{referencefig}. (b) The uncertainty coming from different choices of impact factors. UgD stands for the unintegrated gluon densities, and
LCWF for the light-cone impact factors, as explained in 
Section~\ref{sec-embedding}.\label{sudak}}
\label{fig-variations}
\end{center}
\end{figure}

Furthermore, we show in Fig.~\ref{fig-variations}.a the effect of changes 
in the choice of parametrisation.
We first show the result of a resummation
of log$^2$ terms, including angular ordering, and choosing the scale of 
$\alpha_S$ as $q^2$. As we have seen before,
the single-logs are opposite to the double logs, so that the cross 
section increases if one includes them.
We also show in the same figure the change coming from choosing $\bk_2^2$
as the scale of $\alpha_S$ in the Sudakov form factor.
One sees that the choice of upper and lower limits in the Sudakov integral,
and that of scale in $\alpha_S$ can lead to much larger effects
than the inclusion of subleading logs and constant terms.

The second uncertainty comes from the impact factor, see 
Fig.~\ref{fig-variations}.b. Several choices of unintegrated gluon 
densities, which differ
only in the choice of parametrisation of soft colour-singlet exchange, 
lead to the shaded band, while the impact
factor based on a simple 3-quark light-cone wave function leads to the 
curve. Although both lead to an acceptable fit
to the CDF data, it seems that the curvature is better reproduced by the 
more sophisticated unintegrated gluon
density. Uncertainties due to possible parametrisations of the soft 
region, and to the choice of form factor, amount to a factor of at least 3.

The parametrisations of the proton form factor used in this study contain 
a soft and a hard part.
The hard part was obtained in \cite{IN2000,IvanovPHD} within the 
$k_t$-factorisation formalism,
which is devised for processes with small-$x$ gluons and not too large 
transverse momenta.
Although central exclusive production takes place precisely in this 
regime, one can, in principle,
improve the treatment of the hard part in the spirit of the CCFM equation 
\cite{CCFM,CCFM2},
and one can imagine that this improvement will shift numerical results.
However, we do not expect it to reduce the spread of different predictions
because it comes mainly from our lack of knowledge of the proton form 
factor in the soft region.

To these uncertainties one must add that on the gap survival probability 
and that on the parametrisation of the splash-out,
so that the overall theoretical uncertainty in the calculation is at 
least a factor 400
between the lowest and the highest estimates.

\subsection{Predictions for the LHC}

\begin{figure}
\begin{center}
\epsfig{file=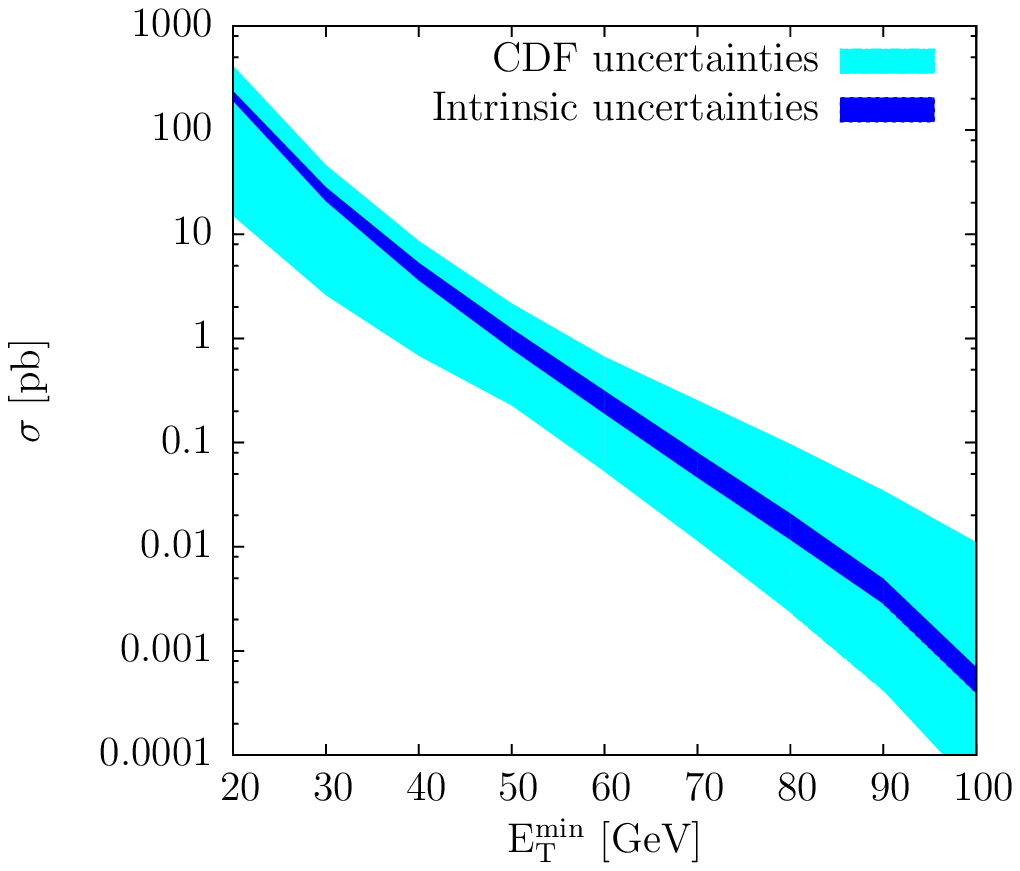,width=8cm}\epsfig{file=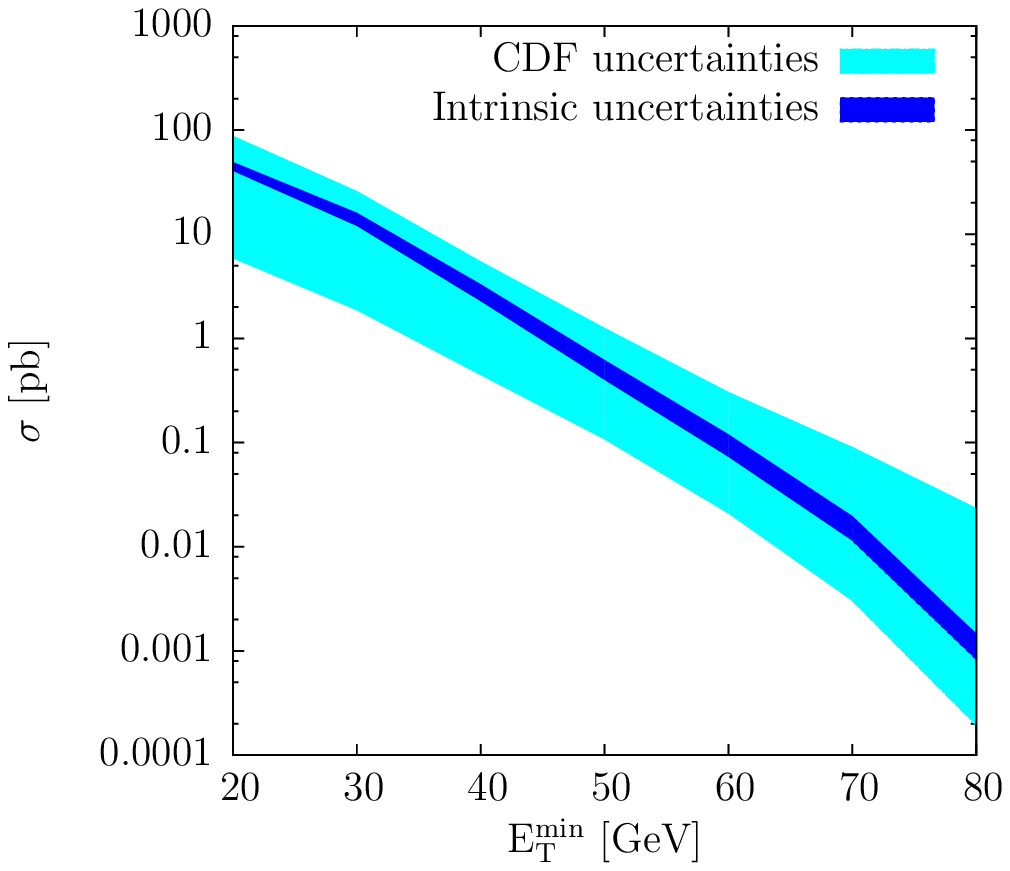,width=8cm}
\caption{The reference curve extrapolated to the LHC, for the two sets of 
cuts shown in Table~\ref{cutsLHC}. The graph to the left corresponds to 
the cuts given in column A and the one to the right to cuts B. The inner 
bands give the theoretical error for
curves reproducing the central values of the CDF data, and the outer 
bands correspond to the 1 $\sigma$ errors.}\label{fig-lhc}
\end{center}
\end{figure}

The formulae derived here can be equally used to estimate the central 
exclusive dijet production at the LHC.
To do this, one needs to take into account the following corrections:
\begin{itemize}
\item
Specific cuts that will be used at the LHC to search for such events. 
These are given in Table~\ref{cutsLHC}.
\item
Extrapolation of the proton form factor to the LHC energies. This can be 
done since
the parametrisations of the unintegrated gluon distributions are 
available for very small $x$.
\item
Extrapolation of the gap survival probability. For an estimate, we will 
use the prescription:
$\langle S^2_{LHC}\rangle=\langle S^2_{Tevatron}\rangle/2$.
\end{itemize}
The predictions for the LHC based solely on theoretical calculations will 
be unavoidably plagued
by the same very large uncertainty as for the Tevatron.
Restricting the models with the Tevatron data considerably reduces this 
uncertainty.
In Fig.~\ref{fig-lhc} we show our predictions for the dijet central 
exclusive production at the LHC.
Shown are two bands indicating the range of uncertainties.
The inner band represents how different sets of parameters, tuned at the 
Tevatron to describe the central value
of the data,
diverge as one extrapolates from the Tevatron to the LHC. The outer band  
includes all the parametrisations presented in the other figures of this 
paper which go through all the CDF points at the $1\sigma$ level.

We also checked that the ratio $\sigma_{pert}/\sigma$ described in
Section~\ref{subsec-properties} is about 0.35 at the LHC,
confirming  that the gluon loop is still dominated by the soft region.
\begin{table}
\begin{center}
\begin{tabular}{|c|c|c|}
\hline  cuts& A \cite{LHC1} &B  \cite{LHC2}\\
\hline  $\alpha_1+\alpha_2$& [0.002,0.02] &[0.005,0.018]\\
\hline  $\beta_1+\beta_2$& [0.002,0.02]&[0.004,0.014]\\
\hline  $|y_{jet}^{(i)}|$& $<1$ &$<1.75$\\
\hline  $|y_{cluster}|={1\over 2}
|\log{\beta_1+\beta_2\over\alpha_1+\alpha_2}|$& $-$ &$<0.06$\\
\hline  $M_{jj}$& $>50$ GeV &$>80$ GeV\\
\hline
\end{tabular}
\end{center}
\caption{Experimental cuts\label{cutsLHC} for the LHC.}
\end{table}

\section{Conclusion}\label{sec-conclusion}
In this paper, we have seen that central exclusive production remains 
dominated by the non-perturbative region.
This means that it is very important to use impact factors --- such as 
the unintegrated gluon densities of
\cite{IN2000,IvanovPHD,INS2005} --- which take the non-perturbative 
region into account.
We have shown that
this dominance of the soft region implies large uncertainties in all the 
ingredients of the
calculation: vertex corrections, impact factors and screening 
corrections. We evaluate the uncertainty as
being a factor 20 up or down, with no theoretically preferred curve.

At present, one can tune a perturbative calculation to the CDF run II
data on dijet exclusive production, and try to use it to predict the 
cross sections for the production of other
systems of particles.

There are several problems with this approach.  The first one is that the 
huge Sudakov form factors of the dijet case
suppress diagrams where the hard scale is concentrated in one propagator 
of the graph. Other graphs, such as those of Fig.~\ref{fig-newsudak},
where the hard scale flows through the diagram, are suppressed by 
propagators, but one may expect the vertex corrections to
be substantially smaller. So  using dijet production as a handle on {\it 
e.g.} Higgs production may be misleading.

The second problem concerns the extrapolation to LHC energy. This is due 
to the fact that unitarisation effects will be important
when going from 2 TeV to 14 TeV. The embedding of the process in a 
multi-pomeron environment can be achieved for
short-distance hard partonic processes, leading to the usual gap survival 
probability. However in general,
if the partons are not at very small distances, the gap survival 
probability will be more complicated, and larger. Also,
it is usually assumed that Regge factorisation holds, i.e. that the 
process can be written as a product of factors depending
on $t_1=k_1^2$ and $t_2=k_3^2$, which is not the case here.

Clearly, all the above questions can only be settled by comparing to more 
data. Hence we believe that a measurement
of quasielatic jet production at the LHC, including the rapidity 
distribution, the mass distribution, as well as the $E_T$
distribution,  will be a very important task that will help to clarify 
many of the issues raised in this paper.

\section*{Acknowledgements} We thank M. Ryskin for communications and 
discussions about
the lowest-order estimate and the Sudakov form factors, V. Khoze for 
pointing out the
importance of the splash-out, and A. Papa for discussions about the 
general structure
of jet production diagrams. We also acknowledge discussions with 
A.~Martin, P.V. Landshoff,
and K. Goulianos, and thank K. Terashi for private communications.

\section*{Appendices}
\appendix
\section{Lowest-order two-gluon production\\
 in the quasi-multi-Regge kinematics}

Consider the production of a colour-singlet two-gluon state in the $qq$
collision. The two gluons are not required to have large
transverse momenta or large invariant mass, but they are well
separated in rapidity from the quarks. Besides, we require as
usual that there is no overall colour flow in the $t$-channel.
Then, the kinematics and all the simplifications discussed in
Sect.~\ref{subsec-kinematics} and \ref{subsec-imaginary} still
hold, and we are left with two generic diagrams shown in
Fig.~\ref{diag2}.

From the BFKL point of view, we deal with two-gluon production in
the quasi-multi-Regge kinematics (qMRK). The imaginary part of the
amplitude can be written as
\bea
\mathrm{Im}{\cal M} & = & \mathrm{Im}{\cal M}_a + \mathrm{Im}{\cal M}_b =
s {g^6 \over 2\pi^2} {\delta^{ab} \over 4N} \int d^2\bk\ 
e^{(1) *}_{\mu_1 a} e^{(2) *}_{\mu_2 b} \label{ampqMRK}\\
&\times&\Biggl[{C_{2}^{\mu_1\mu_2} \over \bk^2 (\bk+\bk_1)^2 (\bk+\bk_3)^2}
 - {C^{\mu_1}{\tilde C}^{\mu_2} \over \bk^2 (\bk+\bk_1)^2 
(\bk+\bk_2)^2 (\bk+\bk_2-\bk_3)^2 }\Biggr]
\,.\nonumber
\eea
Here $C^{\mu_1}$ and ${\tilde C}^{\mu_2}$ are the
usual Lipatov effective vertices (the tilde indicates that this vertex
originates from the other $t$-channel leg), {\it e.g.}
\be
C^{\mu_1} = p^{\mu_1} \left(\alpha_1 - 2 {(\bk+\bk_2)^2 \over \beta_1 s}\right)
- q^{\mu_1} \left(\beta_1 - 2 {(\bk+\bk_1)^2 \over \alpha_1 s}\right)
+ (2\bk+\bk_1 + \bk_2)^{\mu_1}\,, \label{lipatov}
\ee
while $C_{2}^{\mu_1\mu_2}$ is the effective
RRGG vertex for two gluon production in qMRK \cite{RRGG}. In the
multi-Regge kinematics (MRK) limit, $\beta_1 \gg \beta_2$, it factorizes 
into usual Lipatov
vertices describing successive production of two MRK gluons:
\be
C_{2}^{\mu_1\mu_2} \approx C^{\mu_1} {1 \over (\bk+\bk_2)^2}\, C^{\mu_2}\,.
\ee
On the other hand, when $|\bk_2|\gg
|\bk_1|,\,|\bk_3|$, one would recover from (\ref{ampqMRK}) the
amplitude (\ref{ampEPA}) described in the main text.

The two terms in (\ref{ampqMRK}) describe the respective
contributions of the two generic diagrams of Fig.~\ref{diag2}. Let
us estimate how the contribution of the second diagram compares
with the first one in the case of large $\bk_2^2$ and, for
simplicity, at $\beta_1 \gg \beta_2$.

The square of the first diagram is
\bea
|\mathrm{Im}{\cal M}_a|^2 &\propto &
\int {d^2\bk \over \bk^2 (\bk+\bk_2)^2} {d^2\bk' \over 
\bk^{\prime 2} (\bk^\prime + \bk_2)^2}\nonumber\\
& \times &{-2 \over (\bk+\bk_1)^2(\bk^\prime+\bk_1)^2}
\left[{(\bk+\bk_1)^2(\bk'+\bk_2)^2
+ (\bk+\bk_2)^2(\bk'+\bk_1)^2 \over (\bk_1-\bk_2)^2} - 
(\bk-\bk')^2 \right]\nonumber\\
& \times & {-2 \over (\bk+\bk_3)^2(\bk^\prime+\bk_3)^2}
\left[{(\bk+\bk_3)^2(\bk'+\bk_2)^2
+ (\bk+\bk_2)^2(\bk'+\bk_3)^2 \over (\bk_3-\bk_2)^2} - 
(\bk-\bk')^2 \right]\nonumber\\
&\approx & {16 \over (\bk_2^2)^2} \int {d^2\bk \over \bk^2}
{d^2\bk' \over \bk^{\prime 2}} \cdot 
{(\bk+\bk_1,\bk^\prime+\bk_1) \over (\bk+\bk_1)^2(\bk^\prime+\bk_1)^2}\,
{(\bk+\bk_3,\bk^\prime+\bk_3) \over (\bk+\bk_3)^2(\bk^\prime+\bk_3)^2}\,.
\eea
For nearly forward
scattering, $|\bk_1|,\, |\bk_3| \ll |\bk|,\, |\bk^\prime|$, so
that the estimate simplifies to
\be
|\mathrm{Im}{\cal M}_a|^2
\propto {1 \over (\bk_2^2)^2} \int {d^2\bk \over (\bk^2)^2}
{d^2\bk' \over (\bk^{\prime 2})^2}\,,\quad \to \quad
\mathrm{Im}{\cal M}_a \propto {1 \over \bk_2^2} \int {d^2\bk \over (\bk^2)^2}\,.
\label{ampAppestimate1}
\ee
A similar estimate for the second diagram gives
\bea
|\mathrm{Im}{\cal M}_b|^2 &\propto & \int {d^2\bk \over \bk^2 (\bk+\bk_2)^2}
{d^2\bk' \over \bk^{\prime 2} (\bk^\prime + \bk_2)^2}\nonumber\\
& \times & {-2 \over (\bk+\bk_1)^2(\bk^\prime+\bk_1)^2}
\left[{(\bk+\bk_1)^2(\bk'+\bk_2)^2 + (\bk+\bk_2)^2(\bk'+\bk_1)^2 
\over (\bk_1-\bk_2)^2} - (\bk-\bk')^2 \right]\nonumber\\
& \times & {-2 \over (\bk+\bk_2-\bk_3)^2(\bk^\prime+\bk_2-\bk_3)^2}
\left[{(\bk+\bk_2-\bk_3)^2\bk^{\prime 2} + \bk^2(\bk'+\bk_2-\bk_3)^2 
\over (\bk_3-\bk_2)^2}\right. \nonumber\\
&&\phantom{ {-2 \over (\bk+\bk_2-\bk_3)^2(\bk^\prime+\bk_2-\bk_3)^2}}
 \left.- (\bk-\bk')^2 \right]\nonumber\\
&\approx & {16 \over (\bk_2^2)^2} \int {d^2\bk \over \bk^2} {d^2\bk' 
\over \bk^{\prime 2}}
\cdot {(\bk+\bk_1,\bk^\prime+\bk_1) \over (\bk+\bk_1)^2(\bk^\prime+\bk_1)^2}\,
{\bk\bk^\prime \over (\bk_2^2)^2}\,,
\eea
which for the forward case becomes
\be
|\mathrm{Im}{\cal M}_b|^2 \propto {1 \over (\bk_2^2)^4} 
\int {d^2\bk \over \bk^2} {d^2\bk' \over \bk^{\prime 2}}\,,\quad
\to \quad \mathrm{Im}{\cal M}_b \propto {1 \over (\bk_2^2)^2} 
\int^{\bk_2^2} {d^2\bk \over \bk^2}\,.
\label{ampAppestimate2}
\ee
Thus, the second diagram is not only suppressed with respect to
(\ref{ampAppestimate1}) by an extra power of $\bk_2^2$, but also
enhanced by an extra logarithm. If the loop corrections to this
diagram can be resummed and presented in a form analogous to the
Sudakov form factor, then the $\bk^2$-integral will be shifted
towards large values and no logarithmic enhancement in this new
suppressing factor will occur.
Hence the new Sudakov form factor may be much larger than the usual one.

\section{Intermediate production of gluons in multi-Regge \\ kinematics as
a source of corrections\\ to central exclusive production}

\begin{figure}
\begin{center}
\epsfig{file=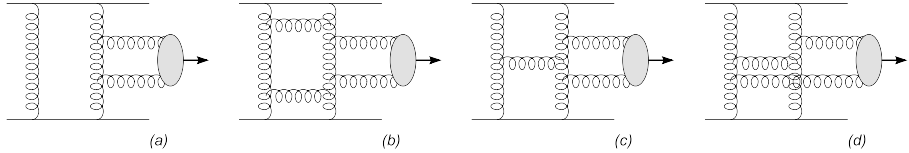,width=12cm}
\caption{{\em a}: Generic diagram representing production of two gluons 
in MRK followed by
their fusion into the high mass system. {\em b}: extra gluon
rungs above or below the central subprocess can be incorporated
into the definition of the unintegrated gluon density.
{\em c}: an extra gluon rung that couples the central subprocess
with the screening $t$-channel gluon represents a single-log
correction to the vertex.
{\em d}: higher-order single-log enhanced correction
to diagram Fig.~\ref{fig-secondcorrection}.c.
}\label{figs-app}
\end{center}
\end{figure}

If a colourless high-mass system $f$ can be ``radiated'' off the 
$t$-channel gluons,
then the same system can be produced via two intermediate gluons in  
multi-Regge kinematics (MRK)
followed by their fusion into $f$, see Fig.~\ref{fig-secondcorrection}.
We do not present here a detailed calculation of this process;
however, it can be immediately
seen from (\ref{dsigma}) and (\ref{ampqMRK}) that the resulting
cross section is flat in rapidities of the intermediate gluons.
Integrating over the corresponding lightcone variables but keeping
$M_f^2$ fixed yields an extra longitudinal logarithm.
The particular diagram shown in Fig.~\ref{figs-app}.a is in fact
a double-log-enhanced correction to the direct production of system $f$.
In the standard calculation it is effectively included via the Sudakov 
form factor.

Even higher-order corrections to this diagram include diagrams with extra 
gluon rungs that
join the two $t$-channel gluons.
If these extra gluons are above or below the central subprocess,
as in Fig.~\ref{figs-app}.b,
they can be absorbed into the definition of the unintegrated gluon density.
However, if the gluon is attached inside the central subprocess, 
Fig.~\ref{figs-app}.c,
it leads to a correction to the effective vertex and cannot be absorbed into
the gluon density.

This correction is enhanced by a single logarithm. If the lightcone variables
of the two intermediate gluons are $\beta_1 \gg \beta_2$, then
the logarithm that appears is of type
$$
\log\left({\beta_1\over\beta_2}\right) \approx 
\log\left({M_f^2\over \bq^2}\right)\,,
$$
where $\bq^2$ is the (moderate) transverse momentum of the intermediate gluons.
Thus, this diagram represents a single-log enhanced correction to the 
central exclusive
production of system $f$, which involves the screening gluons and which 
is absent
in the pure $gg \to f$ process.
Since the single-log enhanced terms in the Sudakov integral are very important
for the kinematic range considered, one must take seriously such corrections.

As the production of two gluons in MRK requires gluon emission from both 
$t$-channel legs,
one encounters additional single-log enhanced diagrams such as shown in 
Fig.~\ref{figs-app}.d.

We would like also to stress that, even without the above estimates,
there is a need to evaluate such ``non-conventional'' diagrams,
which stems directly from self-consistency of one's approach.
Indeed, if one believes the diagrams such as Fig.~\ref{figs-app}.d
are not logarithmically enhanced, then the bare diagram 
Fig.~\ref{fig-secondcorrection}.c
does not get suppressed at all. It can then become comparable with the 
usual Sudakov-suppressed
diagrams, as was explained in Section~\ref{subsec-screening}.
If, on the other hand, Fig.~\ref{figs-app}.d and similar diagrams are 
logarithmically enhanced,
one needs to resum them to estimate their effect.

\end{document}